\def\useieeelayout{1}
\def\showall{0}

\newcommand{\inConf}[1]{\if\showall1{\color{green!50!black}In Journal: #1}\else\if\useieeelayout1{#1}\fi\fi}

\newcommand{\inArxiv}[1]{\if\showall1{\color{blue}In ArXiV: #1}\else\if\useieeelayout0{#1}\fi\fi}

\if\useieeelayout1
\documentclass[letterpaper, 10 pt, conference]{ieeeconf}  
\IEEEoverridecommandlockouts                              %
\usepackage{graphicx}
\usepackage{amsmath}
\usepackage{optidef}
 
\usepackage{amsthm}
\usepackage{amssymb}
\usepackage{amsfonts}       
\usepackage{multicol}
\usepackage{cuted}
\usepackage{multirow}
\usepackage{caption, subcaption}
\newtheorem{theorem}{Theorem}[section]
\newtheorem{lemma}[theorem]{Lemma}
\newtheorem{proposition}[theorem]{Proposition}
\newtheorem{definition}{Definition}[section]
\newtheorem{problem}{Problem}[section]
\newtheorem{assumption}{Assumption}[section]
\newtheorem{remark}{Remark}[section]
\newtheorem{question}{Question}[section]
\newtheorem{conjecture}[theorem]{Conjecture}
\newtheorem{corollary}{Corollary}[theorem]
\usepackage{soul}
\usepackage{cite}
\usepackage[colorlinks=true,allcolors=blue]{hyperref}
\usepackage{comment}
\usepackage{algorithm}
\usepackage{algorithmic}
\usepackage{scalerel}
\usepackage{bm}
\usepackage{multirow}
\usepackage{rotating}
\usepackage{pifont}
\newcommand{\cmark}{\ding{51}}%
\newcommand{\xmark}{\ding{55}}%

\usepackage{textcomp}
\usepackage{multirow}
\usepackage{orcidlink}
\usepackage[intoc, english]{nomencl}
\makenomenclature

\title{\LARGE \bf
A Phased Development Framework Enabling Islanded Operation of Sustainable AI Data Centers With Onsite Grid-Following and Grid-Forming Energy Architectures

}

\author{Soham Ghosh ~\orcidlink{0000-0002-6151-8183}, Nabil Mohammed ~\orcidlink{0000-0002-5068-440X}, and Mohammad Ashraf Hossain Sadi ~\orcidlink{0000-0002-4192-6796}
\thanks{Soham Ghosh is with Black \& Veatch, Overland Park, KS 66211, USA. 
Nabil Mohammed is with the Centre for New Energy Transition Research (CfNETR) at Federation University, Ballarat, Australia. 
Mohammad Ashraf Hossain Sadi is with the University of Central Missouri, Warrensburg, MO 64093 USA.  Correspondence email: sghosh27@ieee.org}%
}

\else
\documentclass[10pt]{article}
\usepackage{graphicx}
\usepackage{amsmath}
\usepackage{optidef}
 
\usepackage{amsthm}
\usepackage{amssymb}
\usepackage{amsfonts}       
\usepackage{multicol}
\usepackage{cuted}
\usepackage{multirow}
\usepackage{caption, subcaption}

\usepackage{xcolor}
\usepackage{cite}
\usepackage[colorlinks=true,allcolors=blue]{hyperref}
\usepackage{comment}
\usepackage{algorithm}
\usepackage{algorithmic}
\usepackage{scalerel}
\usepackage{bm}
\usepackage{url}
\usepackage{multirow}
\usepackage{orcidlink}
\usepackage{rotating}
\usepackage{pifont}
\newcommand{\cmark}{\ding{51}}%
\newcommand{\xmark}{\ding{55}}%

\usepackage{textcomp}
\usepackage{multirow}
\definecolor{steelblue}{RGB}{70,130,180}

\usepackage[preprint]{tmlr}
\usepackage[intoc, english]{nomencl}
\makenomenclature

\title{A Phased Development Framework Enabling Islanded Operation of Sustainable AI Data Centers With Onsite Grid-Following and Grid-Forming Energy Architectures
}

\author{\name Author 1 \email author1@gmail.com \\
      \addr Department of Automatic Control \\ XXX University, Country
      \AND
      \name Author 2 \email author2@gmail.com \\
      \addr Department of Automatic Control\\ XXX University, Country
      \AND
      \name Author 3 \email author3@gmail.com \\
      \addr Department of Automatic Control\\ YYY University, Country
      }

\def\month{MM}  
\def\year{YYYY} 
\def\openreview{\url{https://openreview.net/forum?id=XXXX}} 

\fi

\begin{document}

\maketitle
\thispagestyle{empty}
\pagestyle{empty}

\begin{abstract}
As hyperscale and colocation AI data centers continue to expand, the electric grid is increasingly required to support large, concentrated loads, with individual facilities ranging from 500 MW to 2 GW. Current projections estimate that approximately 50 GW of AI data center capacity will require grid connectivity in the United States by 2030. While prior research has extensively examined the environmental and operational impacts of AI data centers, as well as their potential role as grid-interactive assets, limited attention has been given to the challenges associated with their scalable deployment through engineering, procurement, and construction (EPC) processes. This manuscript addresses this gap by proposing a phased development framework for AI data center expansion. The approach is designed to enable developers to meet aggressive time-to-market objectives while navigating multi-year constraints associated with interconnection approvals and lead times associated with the procurement of component equipment. A modular construction architecture is presented, along with a detailed analysis of integrated energy systems and the role of hybrid on-site generation in supporting incremental capacity growth. Electromagnetic transient simulations (EMT) are used to evaluate system performance, demonstrating that a combination of on-site natural gas generation and grid-forming energy storage can reliably support data center operations during early and intermediate deployment phases. The study further examines the transition to full grid interconnection, including the capability of the data center to operate in islanded mode during grid disturbances. Finally, the manuscript compares grid-forming control strategies for system reconnection and restoration under varying  conditions. 

\end{abstract}


\nomenclature{VRFB}{Vanadium Redox Flow Battery}
\nomenclature{LCO}{Lithium Cobalt Oxide battery}
\nomenclature{LMO}{Lithium Manganese Oxide battery}
\nomenclature{LFP}{Lithium Iron Phosphate battery}
\nomenclature{LNMC}{Lithium Nickle Manganese Cobalt battery}
\nomenclature{LTO}{Lithium Titanate Oxide battery}
\nomenclature{PCC}{Point of Common Coupling}
\nomenclature{PESC}{Power Electronics-based Supercapacitors}
\nomenclature{SMES}{Superconducting Magnetic Energy Storage}
\nomenclature{BART}{Bayesian Additive Regression Trees}
\nomenclature{ATC}{American Transmission Company}
\nomenclature{AESO}{Alberta Electric
System Operator}
\nomenclature{EPRI}{Electric Power Research Institute (EPRI)}
\nomenclature{AI}{Artificial Intelligence}
\nomenclature{GFM}{Grid-Forming (inverter based control)}
\nomenclature{GFL}{Grid-Following (inverter based control)}
\nomenclature{NERC}{North American Electric Reliability Corporation}
\nomenclature{IR}{Interconnection Request}
\nomenclature{BESS}{Battery Energy Storage System}
\nomenclature{ONAN}{Oil Natural, Air Natural (means of transformer cooling)}
\nomenclature{LLM}{Large Language Model}
\nomenclature{VSG}{Virtual Synchronous Generator}
\nomenclature{CGVSG}{Compensated Generalized VSG}
\nomenclature{AVSG}{Adaptive VSG}
\nomenclature{FERC}{Federal Energy Regulatory Commission}
\nomenclature{dVOC}{dispatchable Virtual Oscillator Control}
\nomenclature{SRF}{Synchronous Reference Frame}
\nomenclature{PLL}{Phase Locked Loop}
\nomenclature{FAC}{Facilities Design, Connections, and Maintenance}
\nomenclature{TSOs}{Transmission System Operators}
\nomenclature{ISOs}{Independent System Operators}
\nomenclature{BPS}{Bulk Power System}
\printnomenclature
\section{Introduction}\label{sec_intro}


Data centers have supported global information technology infrastructure for roughly the past three decades, primarily enabling cloud storage and enterprise web and application hosting. Since around 2017, the rapid growth of generative artificial intelligence \cite{sengar2025generative, ooi2025potential}, including large language models \cite{zhao2026survey} and multimodal systems \cite{xie2024large}, has significantly reshaped the industry. This shift accelerated further after 2025 with the emergence of agent-based AI systems \cite{bessa2026current}. To meet these new computational demands, data center design has moved away from traditional CPU-focused architectures, which typically operated at about 7 to 20 kW per rack, toward GPU and TPU-dominated systems that can exceed 100 kW per rack \cite{vlachos2026making}. This increase in power density has also driven the adoption of advanced cooling approaches such as direct-to-chip liquid cooling and full immersion cooling \cite{ruichek2025immersion}. \\

\begin{figure*}[!htbp]
  \centering
  \includegraphics[width=1.0\textwidth]{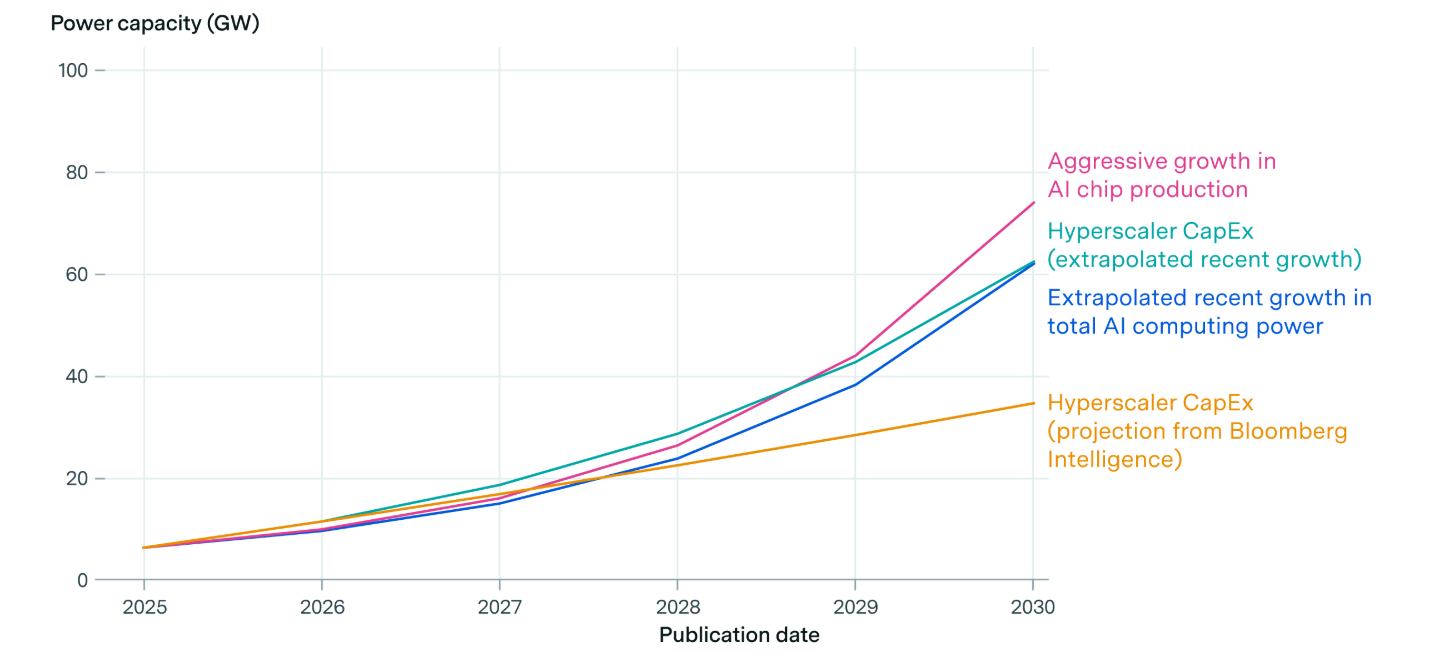}
    \caption{Forecasted capacity of US AI data centers under various project scenarios. Source: epoch.ai. \cite{epoch_ai_dataUSproj}}
    \label{fig:forecasted-total-capacity_US}
\end{figure*}

Looking ahead, a key question is how power requirements will evolve by 2030 and beyond, particularly for training large-scale frontier AI models. Recent EPRI projections \cite{epri2025} suggest that power demand for such training workloads could grow by approximately 2.2 to 2.9 times per year. If this trend continues, individual training runs may require between 4 and 16 GW by 2030. The lower end of this range aligns with already announced plans for multi-gigawatt data center developments. At a global level, total installed AI data center capacity could exceed 100 GW by 2030, with the United States alone potentially accounting for more than 50 GW; see Fig.~\ref{fig:forecasted-total-capacity_US} for forecasted capacity of US AI data centers. This would represent close to 10 percent of the country’s (US) total power generation capacity. As a result, the electric grid is projected to experience unprecedented pressure in terms of growth and power demand, compounded by challenges related to interconnection processes, permitting timelines, equipment procurement, and large-scale construction constraints.

As evident from current market trends, AI data centers have attracted significant attention from both industry and academic research communities in the United States and worldwide \cite{davenport2024ai}. Key areas of focus include intelligent systems for managing computing workloads \cite{padur2022intelligent, gupta2024evolution, sankar2023ai}, advances in optical and photonic communication technologies \cite{yoo2023new}, the development of next-generation memory and high-speed interconnects \cite{vlachos2026making}, and the use of sustainable materials and circular economy practices. In terms of AI data center research that concerns power system engineers, the research horizon is a convergence of multiple focus areas, as outlined in Table~\ref{tab:literature_summary_and contributions}. While studies in terms of the environmental and operational impacts of AI data centers \cite{nahid2025rising, d2024carbon, de2026carbon,farfan2023gone} and AI data centers as grid-interactive assets \cite{colangelo2026ai, lin2024exploding, zhang2025mitigating, zheng2020mitigating} have been carried out, there is a lack of systematic and structured coverage in the area of scalable expansion of AI data centers through large-scale engineering,
procurement, and construction. This manuscript aims to bridge the critical gap. 

\begin{table*}[t]
\small
\centering
\caption{Comparison of prominent related work in the field of AI data center engineering from a power systems standpoint alongside the unique contributions of this manuscript.}
\label{tab:literature_summary_and contributions}
\renewcommand{\arraystretch}{1.2} 
\begin{tabular}{p{2cm} p{10cm} p{4cm}  }

\hline
\textbf{Paper ID} &
\textbf{Factors studies}&
\textbf{Depth of coverage in the literature and other notes} \\
\hline
\cite{nahid2025rising, d2024carbon, de2026carbon,farfan2023gone} &
\textit{Environmental and operational impacts of AI data centers} 

\cite{nahid2025rising} provides region-based market shares of data centers, provides a high-level summary of projected demand, and generation trends 

\cite{d2024carbon} provides a first-of-a-kind life cycle assessment of carbon footprint for an AI datacenter, emphasizing nuclear power (in a hybridized setting) as a potential option

\cite{de2026carbon} is a seminal work highlighting the lack of distinction between AI and non-AI workloads in the environmental reports for data center impact assessments. Estimates 32.6 and 79.7 million tons of CO2 emissions in 2025 along with a water footprint reaching 312.5–764.6 billion L

\cite{farfan2023gone} provides cumulative estimates for the European region. Data usage in OECD-Europe is expected to grow from 86 to 225 EB by 2030. Energy consumption of data services is expected to grow from 30 to 113 TWh by 2030.
&
Covered to an extensive degree

Consensus converges on the point that there is a general lack of transparency in data center operators' reporting of energy and water usage. The available research data is generally approximate in nature. \\
\hline

\cite{colangelo2026ai, lin2024exploding, zhang2025mitigating, zheng2020mitigating} &
\textit{AI data centers as grid-interactive assets}

\cite{colangelo2026ai} demonstrates demand response capability of a hyperscale cloud facility through reduction of power usage by 25\% for 3 hours while maintaining service guarantees 

\cite{lin2024exploding} advocates that relaxing new data center reliability guarantees increases the power available to 1.6x–4.1x while maintaining 99.6\% actual power availability for the new DCs, sufficient for the 5-year AI demand

\cite{zhang2025mitigating} provides a theoretical demonstration of mitigating power grid impact from proactive data center workload shifts through a coordinated scheduling strategy that encompasses synergistic traffic, data, and power networks  

\cite{zheng2020mitigating} provides a working framework for reducing renewable energy curtailment through load migration between data centers &

Covered to a moderate level of degree

Literature covers GPU workload 
modulations in real-time based on grid signals to provide demand response without hardware changes or energy storage by throttling latency-tolerant training and inference tasks while preserving critical workload performance.

\\
\hline

\cite{li2024unseen, ahrabi2025ai, ruan2025data, ross2026using} &
\textit{AI data center related power quality issues}

\cite{li2024unseen} focuses on AI training-induced electrical transients

\cite{ahrabi2025ai} summarizes factors influencing data center siting, including power availability, power quality, and regulatory incentives 

\cite{ruan2025data} advocates investigating the impedance characteristics and outlines cooperative mechanisms to alarm and avoid sub-synchronous resonance (SSO) prone situations 

\cite{ross2026using} investigates oscillation suppression through grid-forming inverters

& 
Covered to a moderate level of degree

Existing literature remains focused on the importance of compliance with IEEE-519 and IEEE 1453 for harmonic distortion and flicker limits. Acknowledges sub-synchronous resonance as a problem.  \\
\hline

\cite{mughees2026ai, mo2025learning} &
\textit{AI data center load observability and forecasting}

\cite{mughees2026ai} proposes a short-term, data center-scale power forecasting framework for AI training and inference using Liquid Neural Networks (LNN) 

\cite{mo2025learning} addresses cross-server power prediction in large data centers and the issue of distribution shifts that pose significant challenges for cross-server power prediction &

Coverage remains sparse

The challenge for grid operators is not to dynamically allocate resources, but to ensure enough generating capacity in GW scale to load demand

\\
\hline

\cite{ghosh2026scalable, bashir2026barrier} &
\textit{AI data centers and power grid colocation/co-design} 

\cite{ghosh2026scalable} highlights partially grid-connected and full on-site generation, along with an initial simple cycle co-located deployment with a phased upgrade to a combined cycle plant 

\cite{bashir2026barrier} argues the need for shifting from implicit coexistence to explicit co-development between compute and power infrastructure while highlighting current technical misalignment  &

Coverage remains sparse
 \\
\hline
Our manuscript &
\textit{Scalable expansion of AI data centers through engineering, procurement, and construction}

Our manuscript highlights (a) phased deployment architecture, (b) the need for energy hybridization, and (c) an understanding of GFM controls specific to supporting
data centers &

Represents a  critical issue not covered previously, to the best of the authors' knowledge, and remains an unexplored frontier \\

\hline
\end{tabular}
\end{table*}

The key contributions of this manuscript are as follows: 
\begin{enumerate}
  \item \textit{phased deployment architecture:} A phased approach for AI data center deployment is presented. This phased approach is directed towards data center developers and takes into account current excessive interconnection bottlenecks in the US and globally, as well as equipment lead times. 
  \item \textit{need for energy hybridization:} Uniqueness of AI training load profile and energy hybridization options supporting initial off-grid deployment for an AI data center is presented. 
  \item \textit{Grid-forming and grid-following interactions in AI data center supply systems:} This study investigates BESSs interfaced through power electronic inverters to power AI data centers, focusing on GFM, GFL, and hybrid GFM–GFL dynamic interactions under variable loading and active power setpoint variations. Furthermore, different GFM control strategies are assessed in terms of small-signal stability, transient performance, and voltage–frequency support capabilities for AI data center applications. 
\end{enumerate}

The rest of the paper is organized as follows: Section \S\ref{sec_Background_on_AI_Centers_Developments} presents a comprehensive overview of current challenges that grid operators and data center developers face with the proliferation of AI data centers. Section \S\ref{sec_phased_development} forms the foundation of a phased development approach that may help AI data center developers navigate current grid bottlenecks, including lengthening grid interconnection studies and evolving regulatory requirements for large loads. The discussion in this section evolves to cover the need for a hybridized energy mix for on-site generation. In section \S\ref{sec:modelling_of_GFM_and_GFL}, the focus shifts to control system architectures for various grid-forming and grid-following inverters, which are fundamental to on-site generation based on battery energy storage. Section \S\ref{sec_simulation_results} where results from multiple on-site generation deployment cases are analyzed. Finally, section \S\ref{sec_conclusion} concludes the manuscript.

\section{Background on AI Centers Developments and Associated Challenges}
\label{sec_Background_on_AI_Centers_Developments}
The challenge of serving large data center loads stems from the fact that these loads have complex electrical equipment, such as GPU-based IT racks and variable frequency drives, with unique energy use patterns. On the part of TSOs/ISOs, meeting such energy demand while ensuring grid stability requires specialized interconnection studies, often accompanied by lengthy transmission upgrades. Utilities, meanwhile, can be caught between the financial incentives of rapidly interconnecting large new loads and their responsibility to rigorously evaluate how these additions will affect system reliability and electricity costs. This challenge is compounded in regions like the United States, where relatively low technical and financial entry requirements allow many projects to enter the interconnection queue. For instance, in ERCOT, the queue has expanded significantly, with a substantial share of post-2026 projects yet to submit required studies to the ISO \cite{quint2025practical}. In the absence of stricter readiness criteria, speculative proposals risk congesting the pipeline. Increasing the entry threshold for large-load interconnection requests can help filter out less viable projects, improving overall certainty and the likelihood of successful interconnections.\\
Another key issue in large-load interconnection is the largely customized nature of current processes, which differ across utilities and jurisdictions and create complications for both service providers and customers. This stems from the traditional “load-serving” approach, where utilities are expected to accommodate demand with minimal technical barriers. While this framework has worked for conventional loads, it is not well suited for large-scale demand. Proper evaluation of factors such as ride-through capability, power quality, load composition, and temporal variability (discussed later) is critical for maintaining system reliability and effective planning. For example, in the United States, the NERC FAC-002-4 standard on Facility Interconnection Studies requires transmission planners and planning coordinators to assess the reliability impacts of new end-use facilities. However, there is still no consistent, industry-wide methodology, leaving utilities to define requirements on a case-by-case basis. \\

\begin{figure*}[!htbp]
  \centering
  \includegraphics[width=0.75\textwidth]{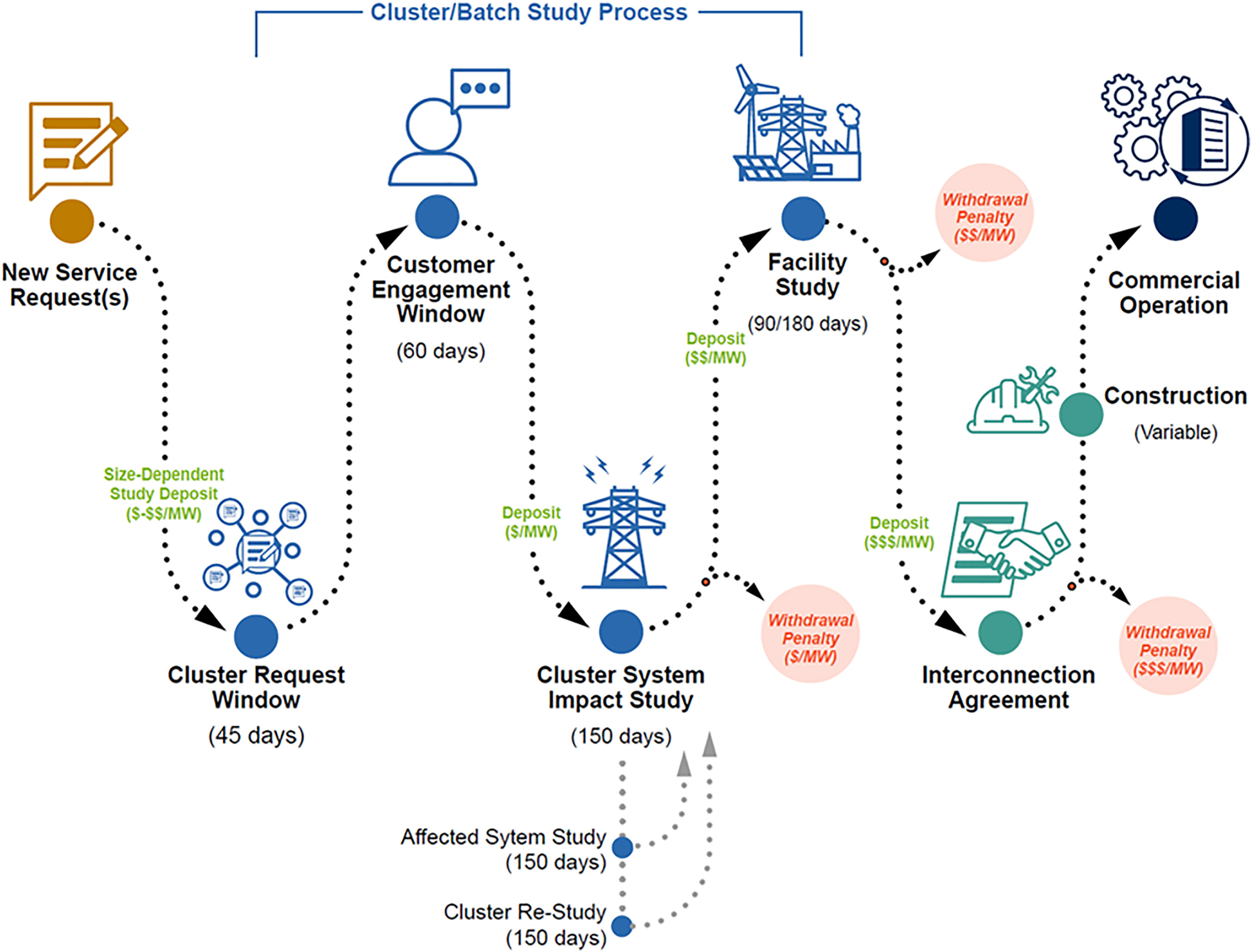}
    \caption{The interconnection process defined under FERC Order 2023 incorporates escalating study deposits and withdrawal penalties as projects advance through successive stages \cite{gorman2025grid}.}
    \label{fig:IA_FERC2023}
\end{figure*}

The industry as a whole is actively working to better navigate the challenges with large loads interconnections. To this end, FERC Order 2023, see Fig.~\ref{fig:IA_FERC2023}, introduces a cluster-based approach to interconnection studies, allowing multiple projects to be evaluated in coordinated batches rather than on a purely sequential basis. This framework helps reduce the need for repeated restudies, which historically arose when projects entered or exited the queue and altered system conditions. In addition, the order standardizes and increases study deposits and withdrawal penalties across regions, creating a more stringent financial commitment as projects progress. As a result, developers face higher costs for late-stage withdrawals, which is intended to discourage speculative entries and improve overall queue discipline.

To better understand the reliability impact(s) of emerging large loads and their impact on the BPS, NERC established the Large Loads Task Force (LLTF) in August 2024, now called the Large Loads Working Group (LLWG); with the NERC working group advancing a three-pronged approach:
\begin{enumerate}
    \item An Essential Action Level 3 alert is being issued (anticipated issue timeframe being May 2026) to recommend near-term mitigation actions that current registered entities can take to address critical reliability risks. 
    \item NERC defining which entities, based on specific physical and electrical criteria, would be required to register with NERC and comply with its reliability standards.
    \item NERC is revising and adopting new reliability standards to define computational loads and establishing measurable requirements for newly registered entities
\end{enumerate}
Being an iterative process, the consolidation of NERC large load action plans and the uniform adoption of applicable reliability standards are expected to continue through Q4 2027 and beyond. Consequently, significant improvements in large-load interconnection processes are unlikely within a 2-3 year timeframe. This underscores the need for a phased development framework to support the ongoing growth trajectory of AI data centers. The following section presents the development of such a framework. 

\section{Proposed methodology for the phased development framework for sustainable AI data centers}
\label{sec_phased_development}
In the previous section, the discussion focused on the emerging trends and interconnection challenges associated with AI data center growth globally and in the United States. The interconnection trends in general and the momentum of large data center loads seeking interconnection to the grid, in particular, are expected to sustain till 2030 and beyond. This sustained momentum in the rapid deployment of AI data centers that rely on grid power, while meeting the siting constraints of permitting, land acquisition, and water-based resources, is extremely challenging to navigate. In this context, this section explores how developers navigating the interconnection bottleneck may benefit from a phased development framework for AI data center deployment and the combination of energy storage systems typically needed for their operation.

\subsection{Phased development of data center plants}\label{sec_phased_development_a}
As seen in section \S\ref{sec_Background_on_AI_Centers_Developments} the wave of interconnection requests into the bulk power system (BPS) is primarily attributed to (a) limited entry barriers, causing a higher degree of speculative load interconnection applications, (b) lack of standardization of the large load interconnection process across jurisdiction, and (c) lack of structured technical requirements such as plant ride-through capabilities, load composition, power quality characteristics, requiring utilities to spend time in tracking missing data essential for these large load interconnection studies. It is against this backdrop that the industry is witnessing a shifting need to a phased development framework for AI data center plants. Such a phased framework is driven by a three-front combination of (a) excessive and growing timeline between the onset of interconnection request (IR) and commencement of commercial operation (COD), (b) ’time-to-deployment’ pressure for AI data center developers, and (c) equipment lead times for off-grid deployment. Table~\ref{tab:lead_times} presents approximate procurement lead times of major equipment needed for an initial off-grid deployment and highlights the fact that certain major equipment, if procured judiciously, may allow the deployment of an initial off-grid AI data center plant within a $\approx$ 24-month time horizon. Against this backdrop, a phased data center construction architecture is introduced in Fig.~\ref{fig:phased_approach}. These three distinct phases are discussed in the following section.\\
\subsubsection{Initial development phase}
Fig.~\ref{fig:phased_approach_a} presents the initial phase with an approximate 24-month deployment time horizon. Within this time horizon, the real estate plot is prepared from a grading, drainage, and erosion control standpoint, along with the development of a substation pad. At this point, the substation consists of a 13.2 kV (or 34.5 kV) bus (will be referred to henceforth as the 'MV bus') with a series of 2000 - 3000 A open-air or switchgear-based circuit breakers connected to it. The choice of open-air breaker design is preferred over metal-enclosed switchgear, given the latter's longer lead times. A certain number of these circuit breakers serve as power feed into data center buildings, while the remainder are interfaced with (a) specialized on-site gas turbine power plants (or gas engines or a combination of both) serving as base power generating unit, and (b) high energy battery energy storage devices connected to the 13.2 kV AC system through inverters and step-up transformers. The justification for this hybrid energy system architecture shall be explored further in section \S\ref{sec_phased_development_c}.   

\subsubsection{Intermediate development phase} Fig.~\ref{fig:phased_approach_b} presents the intermediate phase with an approximate 36-month deployment time horizon. Within this time horizon, the substation is further built with the installation of the main power transformer (shown as a 345/13.2 kV 300 MVA ONAN unit) and the high-voltage circuit breaker (shown as a 345 kV 3000 A unit). This time horizon is also utilized to conduct a series of commissioning tests on the main power transformer and HV circuit breaker. The electrical topology of the 13.2 kV system remains unchanged. Grid connection remains unavailable in this 34-month initial deployment, given the prolonged time horizon for fulfillment of large-load interconnection agreements.

\subsubsection{Final development phase} 
\begin{figure*}[htbp]
    \centering

    \subfloat[Initial islanded built with BESS plus natural gas power turbines without any grid power. \label{fig:phased_approach_a}]{%
        \includegraphics[width=0.80\linewidth]{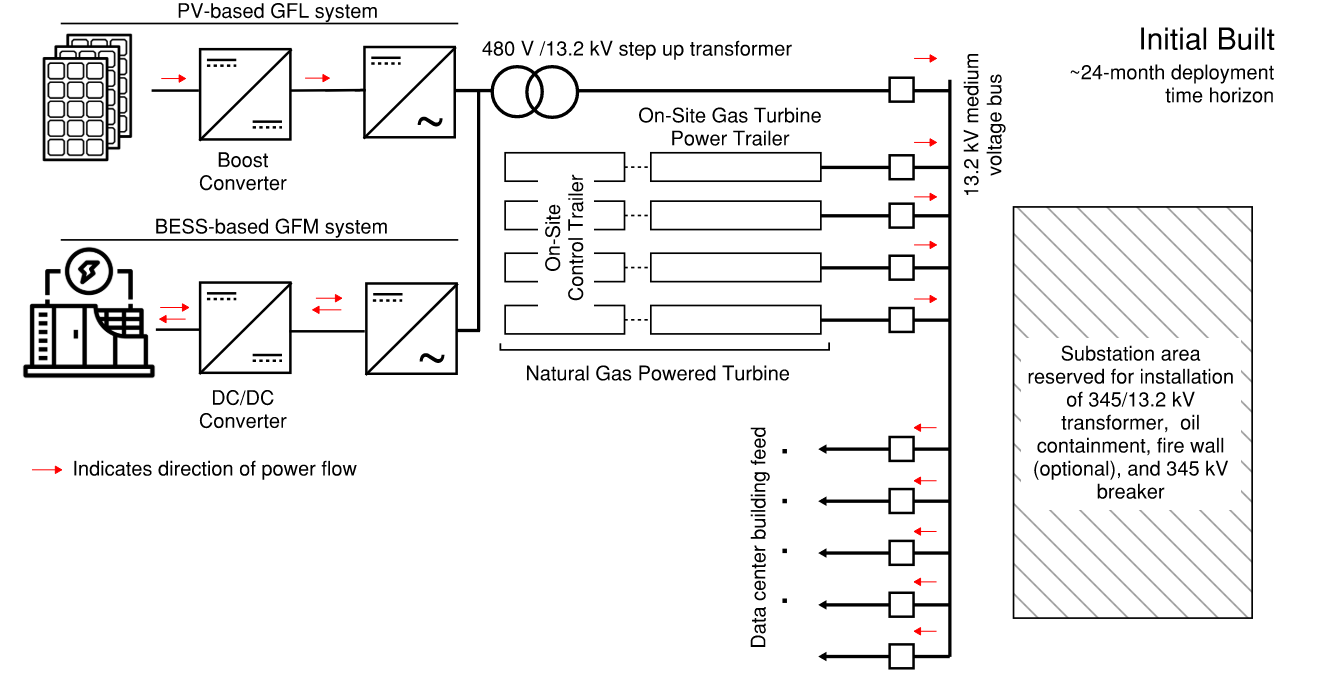}
    }
    \hfill

    \subfloat[An intermediate islanded built showing the placement of large power transformer(s) but without any grid power. \label{fig:phased_approach_b}]{%
        \includegraphics[width=0.80\linewidth]{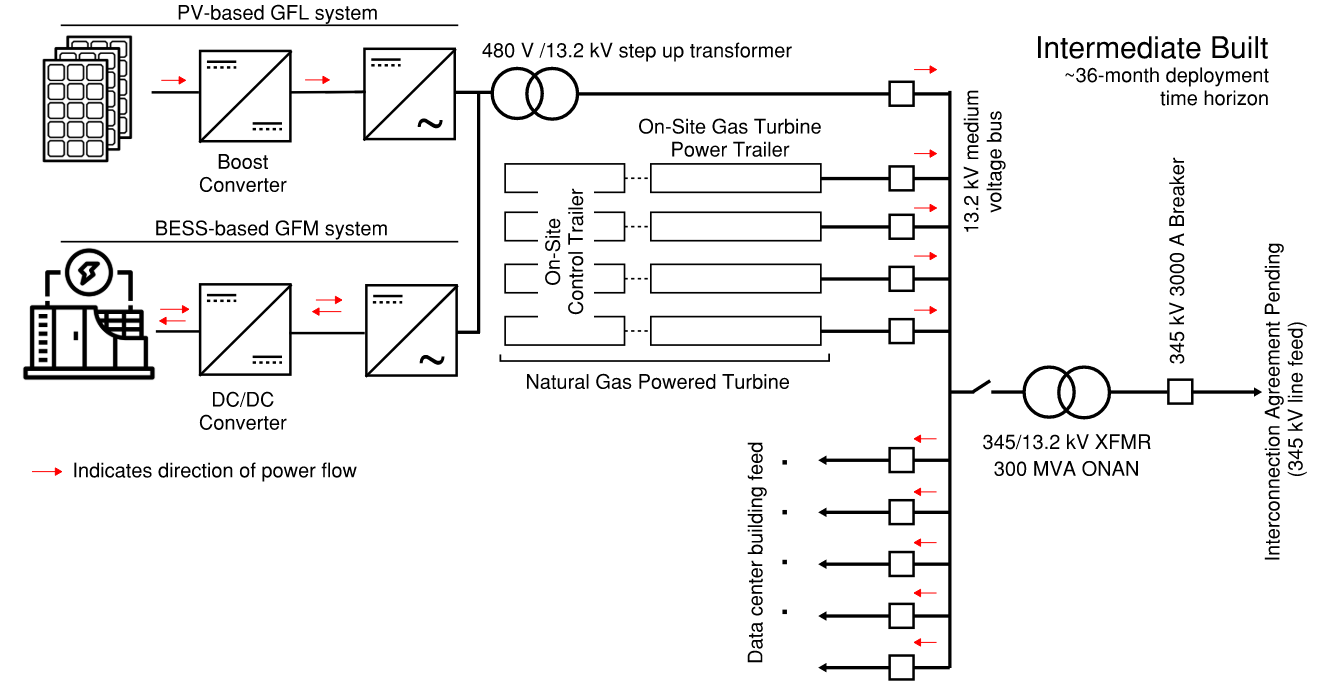}
    }
    \hfill

    \subfloat[Commencement of commercial operation with primary feed from the grid. A portion of the on-site natural gas-powered turbines has been retired, with the feeder laterals used to power additional data center buildings. BESS is retained in service. \label{fig:phased_approach_c}]{%
        \includegraphics[width=0.80\linewidth]{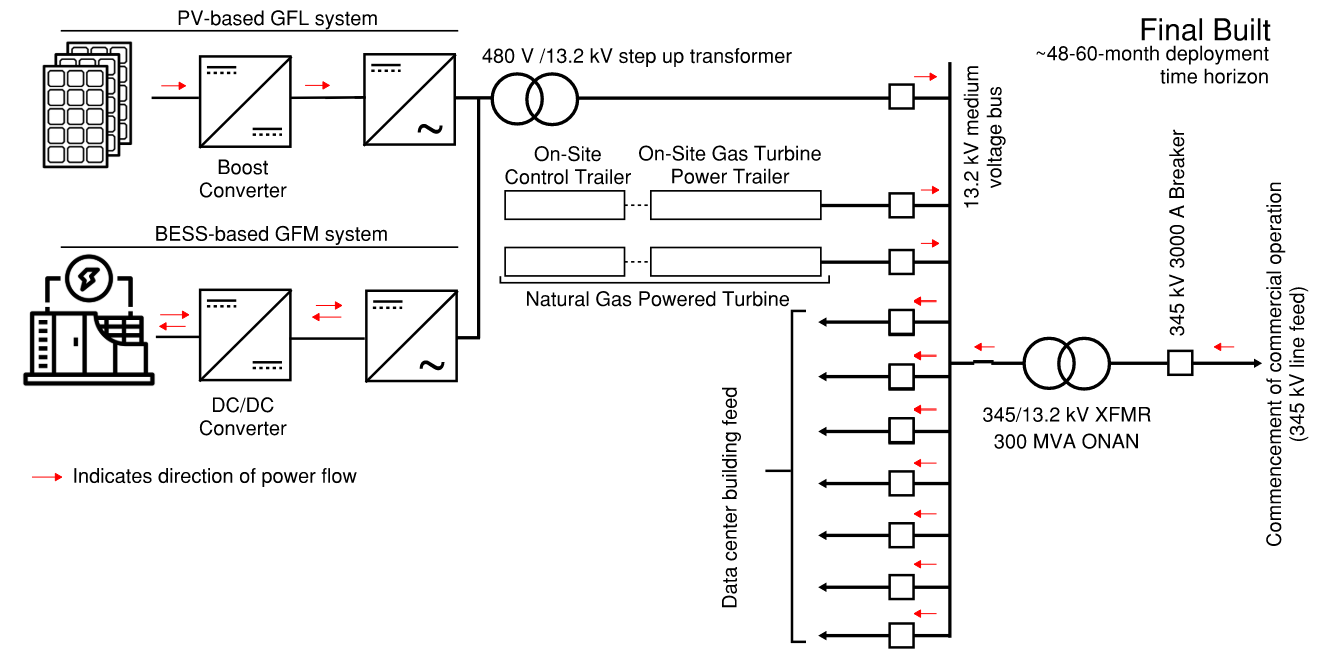}
    }
    \hfill

    \caption{Proposed phased development framework for AI data center construction with initial grid interconnection constraints.}

    \label{fig:phased_approach}
    
\end{figure*}

Fig.~\ref{fig:phased_approach_c} presents the final phase of the development with an approximate 48-60 month deployment time horizon. This phase represents the commencement of commercial operation, with incoming power feed from the grid.  A smaller fleet of on-site gas turbine power plants (or gas engines or a combination of both) is usually retained as an emergency backup power source in the event the plant is forced to operate in an islanded mode due to grid-induced disturbances. The additional 13.2 kV circuit breakers freed up by the reduction in on-site gas turbine power plants are used as additional data center building feeds. In this phase, additional BESS-PV systems may be commissioned and brought online, so that the combined output of the BESS-PV systems and the reduced on-site gas turbine fleet can serve the data center plant loads in the event of a plant islanding. \\
Now that an understanding of the phased development approach for AI data center plants is formed, the discussion shall now focus on forming an understanding of the different stages of AI computation load patterns, while highlighting the particular challenges with AI "training" load patterns and AI "training" data centers. 
 
\begin{table}[t]
\small
\centering
\caption{On-site AI data center generation and substation equipment approximate lead-time}
\label{tab:lead_times}
\renewcommand{\arraystretch}{1.2} 
\begin{tabular}{p{3.5cm} p{2cm} p{1.75cm}  }
\hline
\textbf{Equipment} &
\textbf{Procurement lead time (months)}&
\textbf{Deployement stage} \\
\hline

Substation steel &
3 - 12  & Initial\\

MV breakers &
4 - 6   & Initial\\

Battery system (BESS) with inbuilt MV inverter &
9 - 15  & Initial\\

Natural gas powered simple cycle turbines: Aero-derivative such as Mitsubishi Power FT8® MOBILEPAC® or Solar Titan 350 &
18 - 24 & Initial\\
\hline

HV disconnect switches &
10 - 18 & Intermediate\\

HV breakers &
12 - 57 & Intermediate\\

Main substation transformers &
36 - 52  & Intermediate\\

\hline
Grid interconnection &
48 - 60  & Final\\
\hline
\end{tabular}
\end{table}

\subsection{Understanding the different stages of AI computation load pattern}\label{sec_phased_development_b}
To determine the energy storage systems required to power an AI data center plant cluster, it is important to understand the AI computational load pattern. These AI computational load patterns usually fall into three categories, with training and inference often deployed in physically separate facilities that utilize distinct hardware architectures optimized for their respective computational demands. The current treatment of the various stages of AI computational load patterns remains fragmented across the literature, and this subsection provides a consolidated perspective. Such consolidation is expected to support the development of AI training load profiles presented in the subsequent 'results and discussion' section (section \S\ref{sec_simulation_results}) of the manuscript.

\subsubsection{Training stage} The training stage is often the most power-intensive stage in an AI development process. During the training stage, the AI model is exposed to vast volumes of labeled or unlabeled data, thereby allowing it to learn patterns, correlations, and structures relevant to the task. Table~\ref{tab:Ai_training} presents a summary of some of the most prominent AI models along with their training power drawn, training hardware, and parameter count. \\

\begin{table*}[t]
\small
\centering
\caption{Trends of top-performing AI models and associated training information \cite{EpochAIModels2025}}
\label{tab:Ai_training}
\renewcommand{\arraystretch}{1.2} 
\begin{tabular}{p{2.5cm} p{2.5cm} p{2.75cm} p{2cm} p{3cm} p{2.5cm}}
\hline

\textbf{Model} &
\textbf{Domain} & 
\textbf{Number of parameters} & 
\textbf{Training compute (FLOP)} & 
\textbf{Training hardware and quantity} &
\textbf{Training power draw (W)}\\
\hline

Grok 3 &
Language, vision, multimodal & 
3,000,000,000,000 &
3.5e+26 & 
NVIDIA H100 SXM5 80GB, 80000 &
109,948,656.2 \\

Llama 4 Behemoth (preview) &
Multimodal, language, vision & 
2,000,000,000,000 &
5.184e+25 & 
NVIDIA H100 SXM5 80GB, 32000 &
43,933,454.9 \\

Gemini 1.0 Ultra &
Multimodal, language, vision & 
- &
5e+25 & 
Google TPU v4, 57000 &
38,423,900.1 \\

Llama 3.1-405B &
Language & 
405,000,000,000 &
3.8e+25 & 
NVIDIA H100 SXM5 80GB, 16384 &
22,622,532.1 \\

GPT-4 (Mar 2023) &
Multimodal, language, vision & 
1,800,000,000,000 &
2.1e+25 & 
NVIDIA A100 SXM4 40 GB, 25000 &
19,944,368.1 \\

GPT-4 (Jun 2023) &
Multimodal, language, vision & 
1,800,000,000,000 &
2.1e+25 & 
NVIDIA A100 SXM4 40 GB, 25000 &
19,944,368.1 \\

Amazon Titan &
Language, image generation & 
200,000,000,000 &
4.8e+24 & 
NVIDIA A100, 13760 &
10,929,327.2 \\

MegaScale (Production) &
Language & 
530,000,000,000 &
3.9e+24 & 
NVIDIA A100, 12288 &
9,728,028.1 \\

\hline
\end{tabular}
\end{table*}

The typical load pattern of the AI training phase involves an initial steep ramp-up phase, and a generally constant high power demand \cite{li2025ai} for a prolonged duration. A distinct characteristic of the AI training load pattern is its rapid and large power swings, resulting from intensive matrix computations and dips during lighter operations like data transfers, or when the training state is saved to storage, typically lasting a few milliseconds \cite{mughees2025short}. Network issues during gradient synchronization across hundreds of thousands of GPUs can also cause GPU compute to idle for up to several seconds. After completing a large training job, which often runs for days or weeks, power can ramp down massively if no immediate workload is queued to utilize the GPUs.\\
As shall be seen, the amplitude of these power swings related to AI training complicates the deployment of standalone on-site generation sources such as simple-cycle natural gas turbines and often warrants the hybridized
energy mix for on-site generation. Effective solutions that may reduce these AI power swings are limited and often require a multi-pronged approach, through (1) software-driven controlled workload injection and adjustments, (2) firmware at the GPU-level that flattens ramp rates and minimum power floors, and (3) rack-level battery or capacitor-based energy storage that absorbs or releases quick bursts of power as needed \cite{choukse2025power, zhao2023sustainable, maheshwari2026ai}.
Considerations on ramp rates are important for large grid-connected hyperscale data centers, with ramp rate criteria varying across jurisdictions. As shown in Table~\ref{tab:ramp_rate}, some jurisdictions have defined ramp rate constraints serving as a safeguard, while other jurisdictions have yet to adopt ramp rate limits in place. In the future, it is expected that task forces such as the NERC Large Loads Working Group (LLWG)
to have a broader and more uniform implementation of ramping limits as they better understand the reliability impact(s) of these emerging large loads. 

\begin{table}[t]
\small
\centering
\caption{Active power ramp rate limits for large load across jurisdictions}
\label{tab:ramp_rate}
\renewcommand{\arraystretch}{1.2} 
\begin{tabular}{p{2.75cm} p{5cm}  }
\hline
\textbf{Transmission operator} &
\textbf{Maximum allowable ramp rate limit (MW/ minute)}\\
\hline
Fingrid - Finland &
Adjustable from 5 - 100 \% of $P_{max/min}$ but with a maximum of 50 MW/min \\
\hline
Energinet - Denmark &
60 MW/min \\
\hline
AESO - Canada &
10 MW/min \\
\hline
ATC - US &

Fast oscillations must remain $<$ 25 MW magnitude for periods $<$ 5 seconds, 

Step changes $>$ 50 MW must ramp at $<$ 0.5 MW/s \\
\hline
Southern Company - US &
Active power ramp rate not to exceed 20 MW/min on a 10-minute average \\ 

\hline
\end{tabular}
\end{table}

\subsubsection{Fine-tuning stage} During the fine-tuning stage, a pre-trained AI model is further trained on a smaller, task‑specific dataset so that its parameters adapt to the target domain while preserving its general knowledge. The process updates selected model weights using gradient-based optimization on this new data, improving performance on the desired task without training a new model from scratch. In the fine-tuning phase, the model still draws substantial power, but the demand tends to come in shorter, intermittent bursts and over a much shorter overall period than a full pre‑training run. This reflects the fact that fine-tuning typically uses fewer optimization steps and smaller task-specific datasets, so high GPU utilization is activated in spikes rather than maintained continuously for weeks or months as in full-scale training \cite{li2024unseen}.\\
During the fine-tuning process, approaches such as low-rank adaptation, quantization, and targeted parameter updates are often employed to reduce compute requirements without sacrificing performance \cite{singhapoo2025fine}. Fine-tuning smaller models like BART-base, rather than full-scale architectures, can achieve competitive performance on many tasks while offering significantly improved energy efficiency and reduced computational cost \cite{jeanquartier2026assessing, rehman2025green}.
\subsubsection{Inference stage} AI inference is the stage where a trained AI model is actually used to generate outputs or predictions from new, unseen data, rather than continuing to learn from examples. It is the “execution” or “doing” phase of AI, where the model applies what it learned during training to solve real-world tasks such as classifying images, answering questions, or detecting fraud. Inference is usually the least energy-intensive phase of the different stages of AI computation. The inference stage, however, is a significant portion of the AI energy demand, given that inference runs on millions of user queries.  The energy usage pattern of the inference stage, in general, has the following characteristics (1) it follows a diurnal load pattern with peak load (90 - 100 \% of rated capacity) during business hours and dropping to a base load (45 - 50\% of rated capacity) usually between 1 AM - 5 AM, (2) the weekly cycle usually repeats with 5 - 10\% variability, and (3) bursty spikes from user activity and task complexity, which can be managed by distributing across geographical areas with relatively low inference-related computation workload. \\

It is worth noting that, to handle the growing demand for real-time inference, specialized serving frameworks have become essential, with modern engines such as vLLM and SGLang \cite{vlachos2026making} specifically designed for high‑throughput LLM inference and serving. Additionally, techniques such as pruning and knowledge distillation can help improve the performance and energy efficiency of the inference stage \cite{chen2025electricity}. Another feature of the inference stage worth highlighting is the fact that not all inference tasks are equally energy-intensive. Reference \cite{luccioni2024power} presents an excellent systematic comparison of various interference tasks and their energy-intensive nature, with text classification being the least energy-intensive, while text-generation, image captioning, and image generation are more energy-intensive, determined as tasks with the highest amount of energy intensity. \\

\begin{figure}[!htbp]
  \centering
  \includegraphics[width=0.5\textwidth]{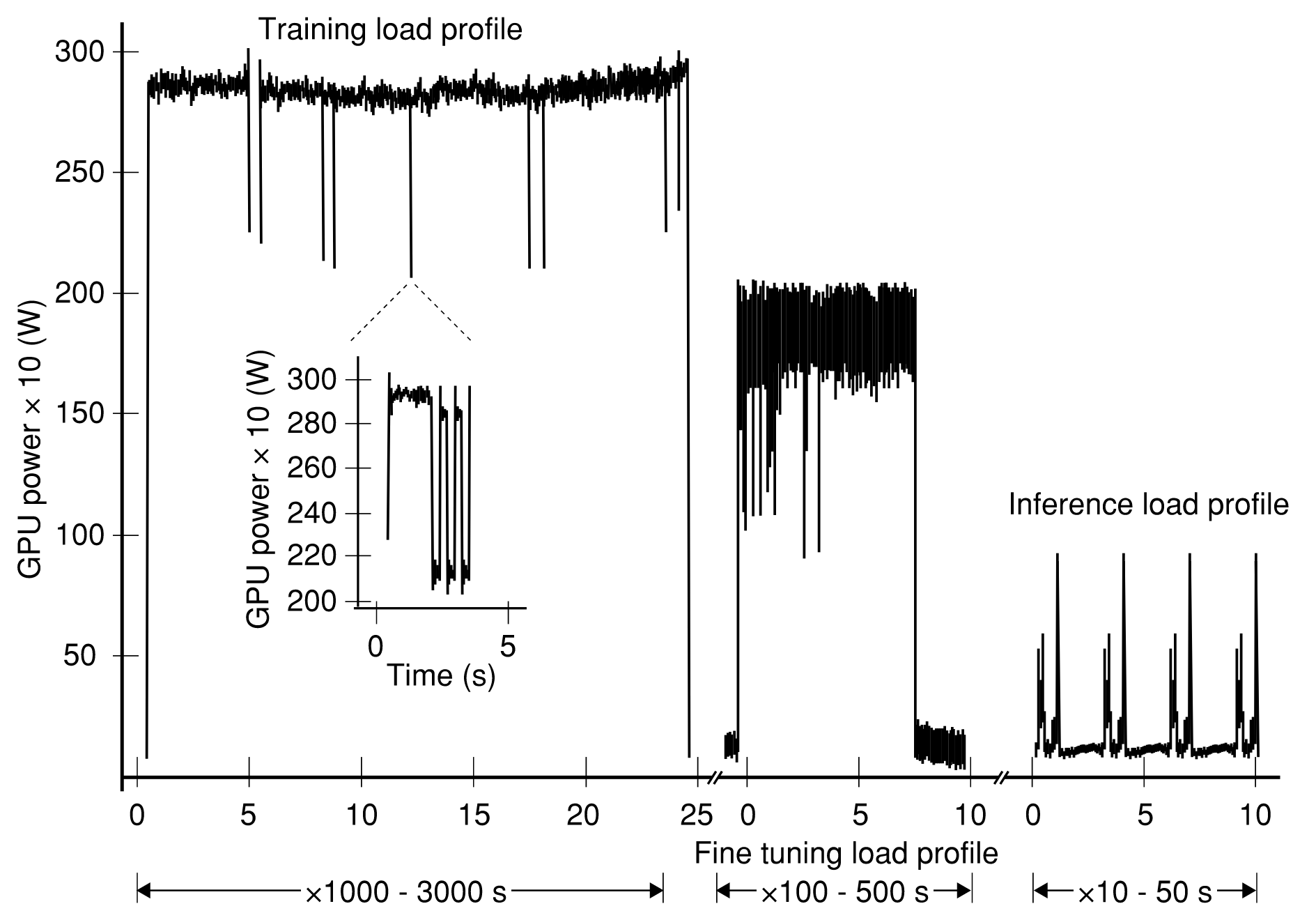}
    \caption{GPU power consumption for different stages of AI computation pattern. }
    \label{fig:Train_ft_inference}
    
\end{figure}

Fig. \ref{fig:Train_ft_inference} illustrates a typical training, fine-tuning, and inference GPU load pattern on a single NVIDIA GB200 machine within a cluster, training on Mistral 7B LLM. The training phase is the longest in duration and the most energy-intensive, and involves the GPUs running continuously for weeks or even months. The training phase exhibits temporal fluctuations arising from synchronizations, checkpoint events, and data loading. A zoomed-in subplot within Fig. \ref{fig:Train_ft_inference} shows multiple temporal fluctuations within a five-second time horizon due to the cumulative effect of such synchronizations and related processes. What is also evident from the preceding discussion is the fact that training (training data centers) creates the foundational AI models that enable all downstream applications. Without trained models, inference cannot occur, making training the essential first step in the AI pipeline. It is for this reason that the focus of this manuscript shall remain on training AI data centers and their phased deployment framework.

\subsection{Energy systems and the need for a hybridized
energy mix for on-site generation} \label{sec_phased_development_c}
This section presents a comprehensive overview of the different energy systems that may form the foundation for on-site generation for AI data centers. 
\subsubsection{Gas turbines and gas engines} Specialized simple-cycle gas turbines fueled by natural gas or flexible low-carbon fuels such as hydrogen, biogas, biomethane/RNG, bio‑propane, or HVO can serve as the primary on-site generation backbone for data centers by providing dispatchable, high-availability power that is decoupled from grid constraints and interconnection delays. Their modular, factory-packaged nature, especially in aeroderivative and mobile configurations, allows 10–40 MW class blocks to be delivered, installed, and commissioned in months or even days, enabling “rapid-deployable” capacity that can closely track phased data hall build‑outs and AI load growth. \\
Such specialized gas turbines, such as Mitsubishi's aero-derivative FT8 MOBILEPAC modular design with a rated output power of $\approx$ 30MW are fast to deploy and come pre-commissioned and can be made site-ready within one day from arrival at the site. Higher output models, such as the SWIFTPAC®60, come as a factory-assembled modular design and can be made site-ready in one month from arrival at the site. Other variants in this class of generators include the Solar Titan 250 model with 27 MW rated output and the Titan 350 model with 38 MW rated output. \\
Though these specialized gas turbines can be deployed much more rapidly and offer parallel operation capability compared to traditional gas turbines in the 100 - 300 MW range, they still come with an 18 - 24 months lead time (see Table~\ref{tab:lead_times}). An emerging solution to this issue is to combine and operate in parallel a set of these specialized gas turbines with reciprocating gas engines such as the Wärtsilä 12V32 (or 18V50) model with output power in the 5 - 18 MW/unit range. An example of such a flexible baseload on-site generation combination may be a set of 6 × Titan 350 gas turbines with 20 × Wärtsilä 12V32 gas engine units, producing a net rated base power output of 340 MW. Table~\ref{tab:gasenginesgasturbines} outlines key characteristics of these AI data center-specific gas turbines and gas engines.

\begin{table}[t]
\small
\centering
\caption{Approximate operational characteristics of gas turbines (GTs) and gas engines (GEs)}
\label{tab:gasenginesgasturbines}
\renewcommand{\arraystretch}{1.2} 
\begin{tabular}{p{3.5cm} p{4cm} }
\hline
\textbf{Characteristics} &
\textbf{Value} \\
\hline
Rated output power range &
$\approx$ 30-60 MW/unit (GTs), 

$\approx$ 5-18 MW/unit (GEs)  \\

\hline
Electrical efficiency &
$\approx$ 45-50\% (GTs), 

$\approx$ 35-40\% (GEs) \\

\hline
Start-up time &
$\approx$ 10 min$^\vartriangle$$^\vartriangle$ (GTs), 

$\approx$ 30 seconds$^\vartriangle$ - 5 min$^\vartriangle$$^\vartriangle$ (GEs) \\

\hline
Ramp rate &
$\approx$ 10 - 15 \% (load/min) (GTs), 

$>$100 \% (load/min) (GEs)\\

\hline
\end{tabular}
\\[3pt]
\footnotesize\textsuperscript{*} Source: authors' compilation based on industry experience\\
$^\vartriangle$ fast-start mode, $^\vartriangle$$^\vartriangle$ cold-start mode
\end{table}
\subsubsection{Flow battery} A flow battery is a rechargeable electrochemical battery in which energy is stored in liquid electrolytes held in external tanks and pumped through a cell stack to charge and discharge. With a flow battery, its power (stack size) and energy (tank volume and electrolyte amount) are physically decoupled, so you can scale MWh by enlarging tanks without redesigning the power block. Because energy and power are decoupled, flow batteries are well‑suited to multi‑hour grid storage applications \cite{jiang2025battery} where one may increase duration by simply scaling tank volume. Recent advancements have been made in membrane-free redox flow battery technology \cite{wang2025recent} while leveraging the utilization of immiscible electrolyte solvents and the engineering of laminar flow dynamics to achieve efficient electrolyte separation without traditional ion-exchange membranes. \\
Though given the recent advancement in the technology and its application in grid energy storage applications, flow batteries generally remain unsuitable for AI data center applications primarily because of their large physical footprint, poor round-trip efficiency, and poor ramping capabilities. However, for green field data center deployments serving cloud-based applications or hosting client applications (enterprise), flow batteries may have some usage.

\subsubsection{Lithium-ion battery} 
\begin{figure}[htbp]
    \centering

    \subfloat[Storage capacity of containerized LFP BESS solution. \label{fig:LFP Trends_A}]{%
        \includegraphics[width=0.9\linewidth]{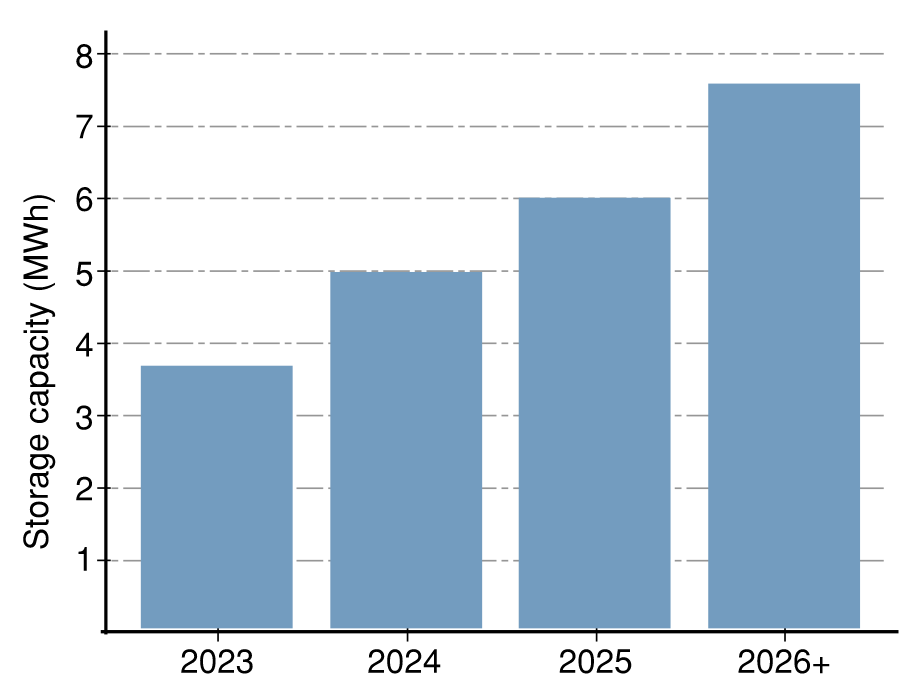}
    }
    \hfill

    \subfloat[LFP BESS price in US dollars. \label{fig:LFP Trends_B}]{%
        \includegraphics[width=0.9\linewidth]{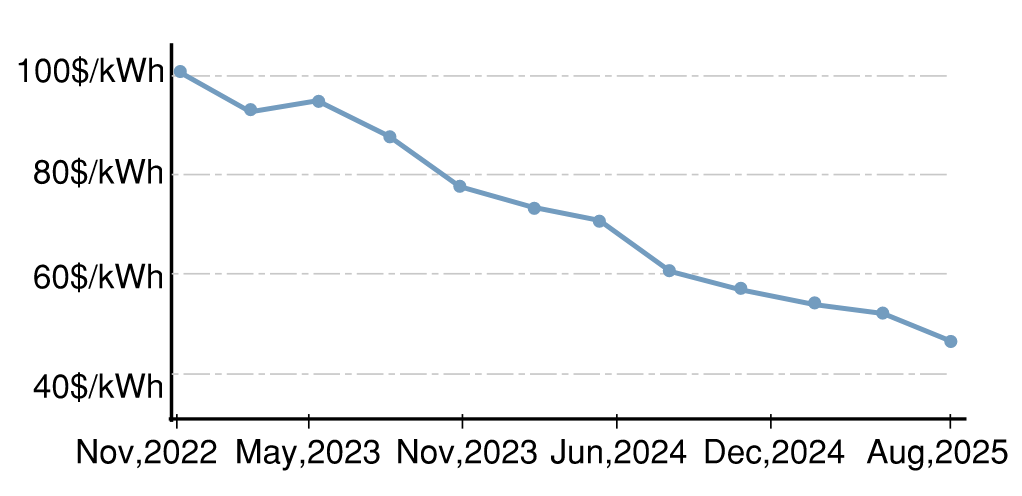}
    }
    \hfill

    \caption{Year-over-year trend for containerized LFP BESS solution.}
    \label{fig:hybrid_energy_sources}
\end{figure}

Dozens of lithium-ion battery chemistries have been explored through research and brought to commercial markets. Table~\ref{tab:hybrid_ESS} summarizes the most common lithium-ion battery chemistries adopted in the utility-scale BESS market and suitable for AI data center applications as a high-energy power source with a response time in the range of 100 ms.
Among the different lithium-ion battery chemistries, lithium iron phosphate (LFP) remains the forerunner in the commercial market with containerized solutions offering storage capacity upwards of 6 MWh/container (see Fig.~\ref{fig:LFP Trends_A}), thereby offering excellent real estate usage, and low relative cost in the range of \$40-60/kWh (see Fig.~\ref{fig:LFP Trends_B}). In addition, LFP batteries sustain very high C-rates, often in the range of 1C- 10C (charge) and up to 25C for discharge, making them particularly suitable for AI data center applications. 

\subsubsection{Superconducting magnetic energy storage (SMES)} An SMES (Superconducting Magnetic Energy Storage) system is an energy storage device that stores electrical energy as the magnetic field created by direct current circulating in a large superconducting coil. Because the conductor is superconducting (essentially zero resistance at cryogenic temperature), the current can circulate with very low loss, so energy can be drawn from or injected into the coil almost instantaneously \cite{vulusala2018application, adetokun2022superconducting}. There are predominantly three different SMES types of control methodologies that are available: (a) thyristor-based control, (b) voltage-source converter based control, and (c) current-source converter based control, with \cite{ali2010overview} serving as an excellent reference for their detailed architecture. For data center applications, the voltage-source converter-based SMES control is usually the preferred choice, as it can control both active and reactive power independently and achieve lower total harmonic distortion compared to thyristor-based SMES control. \\
The cost of an SMES system comes as two components: (a) energy storage related cost, which includes the coil structure, the cryogenic vessel, and the cooling refrigeration, (b) power-conversion related cost. According to \cite{karasik1999smes}, the energy storage related cost ranges between \$85,000 - 125,000/MJ while the power-conversion related cost ranges between \$150 - 250/kW, depending on the control architecture. While SMES systems provide excellent ramp rates and efficiency in the range of 95 - 99\%, also see Table~\ref{tab:hybrid_ESS}, their energy capacity is limited, and the cryogenic complexity from an installation and commissioning standpoint is very high. Compounding that problem, the vendors offering commercial-scale SMEs are fairly limited, with procurement lead times ranging between 36 and 60 months. With 'time-to-deployment' being a primary driver for data center developers, such procurement lead times and engineering complexity make SMESs generally unsuitable for current data center applications. 
\subsubsection{Power electronics-based supercapacitors (PESC)} The last form of energy storage system that shall be discussed in this section is power electronics-based supercapacitors. Supercapacitors enable rapid charge and discharge cycles, with the ability to inject or absorb large bursts ($\approx$ 10,000 W/kg) of active power within milliseconds, with a power conditioning unit often in the form of a modular multilevel converter (MMC) \cite{perez2021modular}. Similar to SMES, PESCs are high-power low-energy-density ($\approx$ 0.2 Wh/L) devices; recall lithium-ion batteries having much higher energy densities of $\approx$ 100 - 270 Wh/kg. \\

Applications of both SMESs and PESCs for AI data centers remain limited to cases where the AI training load that fluctuates orders of magnitude faster than the response time of lithium-ion battery-based storage solutions ($\approx$ 100 ms). Under these scenarios, SMESs and PESCs may absorb sudden GPU power spikes, protecting the lithium-ion battery-based storage solutions from transient overloads \cite{rahman2026energy}. Table~\ref{tab:hybrid_ESS} summarizes the characteristics of common energy storage systems in terms of their power rating, energy density, response time, round-trip efficiency, and ramp rates.\\

\begin{figure*}[htbp]
    \centering

    \subfloat[Hybrid energy systems built with synchronous machines for base power, BESS or flow battery services as high energy sources, and E-STATCOM or SMES serving as high power sources. \label{fig:hybrid_energy_sources_a}]{%
        \includegraphics[width=0.90\linewidth]{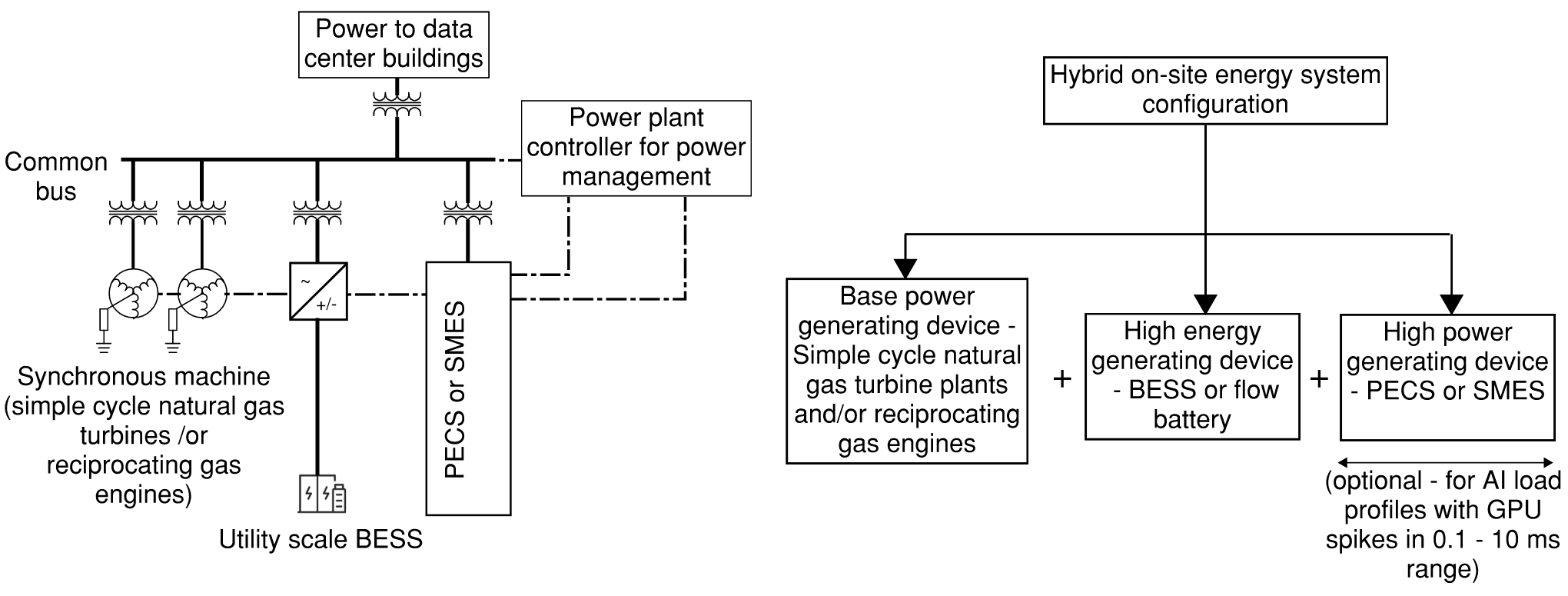}
    }
    \hfill

    \subfloat[Plot showing ramping rates of generating sources including on-site natural gas turbines, lithium-ion based battery system, and SMESs or PESCs. \label{fig:hybrid_energy_sources_b}]{%
        \includegraphics[width=0.90\linewidth]{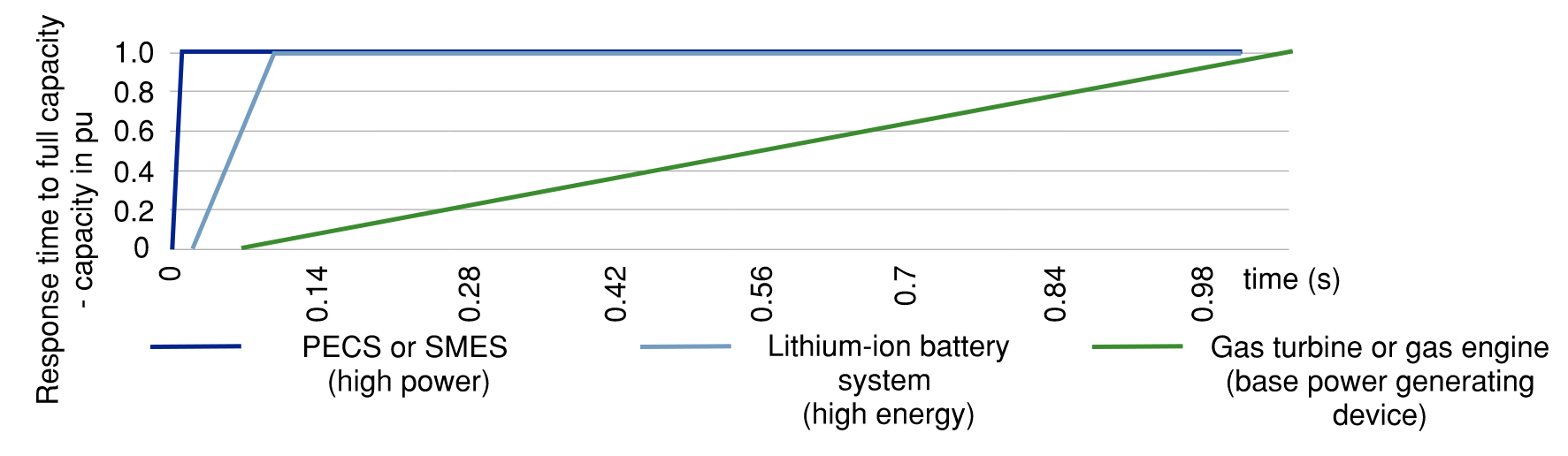}
    }
    \hfill

    \caption{Hybrid energy system combinations for on-site AI data center deployment.}
    \label{fig:hybrid_energy_sources}
\end{figure*}

As evident from the preceding discussion, islanded AI data center deployment requires careful consideration of the different energy sources to manage the power variation associated with AI training load profiles. Fig.~\ref{fig:hybrid_energy_sources_a} shows a hybrid energy system combination with base load generation through simple cycle natural gas-based turbines, combined with a high energy source such as lithium-ion batteries with response time in the range of 100 ms, suitable for most AI training applications, and high-powered devices such as SMESs or PESCs required for very specific AI training applications. Fig.~\ref{fig:hybrid_energy_sources_b} illustrates the ramping rates of base power devices such as simple cycle natural gas-based turbines, with ramp rates in the 1 - 10 second range, high-energy devices such as lithium-ion batteries with ramp rates in the 100 ms range, and high-powered devices such as SMESs and PESCs with ramp rates in 0.1 - 10 ms range. \\
It must be noted here that though SMESs and PESCs serve as high-power storage devices, and can absorb sudden GPU power spike, their deployment  remains marginalized primarily due to two factors:
\begin{enumerate}
  \item GPU power smoothing strategies exist  that allow AI training data center operators to enforce predefined GPU power profiles, with explicit constraints on minimum power floor, ramp-up and ramp-down rates, which allows the operator to limit the on-site generating sources to natural gas turbines and a lithium-ion-based battery system. An excellent discussion of such GPU-based power smoothing, with micro-benchmarking on an NVIDIA GB200 system, may be found here: \cite{choukse2025power}. \\
  
  \item Excessive procurement lead times averaging between 36 and 60 months, which severely restricts their deployment; recall that ’time-to-deployment’ is the primary driver for AI data center developers. 
\end{enumerate}

It is for this very reason that the phased development framework for sustainable AI data center deployment, outlined in section \S\ref{sec_phased_development}, and the simulations and results outlined in section \S\ref{sec_simulation_results}, remain focused on a hybrid energy system comprising natural gas turbines and lithium-ion-based battery system. Thus far, the discussion has addressed the background of AI data center deployments and the associated grid integration challenges, along with a proposed phased deployment framework that accounts for major equipment lead times for a feasible and phased on-site generation deployment with the option to eventually connect the islanded system to the power grid. An examination of the various stages of AI computational load patterns has also been presented, including the specific challenges associated with AI training loads and the energy systems capable of supporting such demands through a hybridized energy mix. Utility-scale lithium-ion battery systems have been identified as a viable hybridization option to complement base power supplied by on-site natural gas turbines. The following section provides a detailed discussion of the control systems governing the inverters interfacing such utility-scale lithium-ion battery systems. 

\begin{sidewaystable*}
\small
\centering
\caption{Characteristic of common energy storage systems (ESS) used in AI data centers \cite{rahman2026energy,adetokun2022superconducting, liu2024review, nazaralizadeh2024battery, nyamathulla2024review, hannan2017review, elalfy2024comprehensive, lebowitz2025}}
\label{tab:hybrid_ESS}
\renewcommand{\arraystretch}{1.2} 
\begin{tabular}{p{3cm} p{2.5cm} p{2cm} p{2cm} p{2cm} p{2cm} p{2cm} p{4cm}}
\hline
\textbf{ESS category} &
\textbf{ESS sub-category} & 
\textbf{Power rating} & 
\textbf{Energy density (Wh/kg)} & 
\textbf{Response time} & 
\textbf{Efficiency (\%)} & 
\textbf{Ramp rate} &
\textbf{Notes}\\
\hline
Flow battery* &
Vanadium redox flow battery (VRFB) &
5 kW - 200 MW &
16 - 35 (Wh/L) &
100 ms & 
~ 70 &
Poor & 
Fairly limited in AI data center application due to poor ramp rate and low round-trip efficiency \\
\hline
\multirow[c]{15}{*}{Lithium-ion battery} &
Lithium cobalt oxide (LCO) &
kW range &
150 - 200  &
100 ms &
90 - 95 &
Limited & 
Susceptible to thermal runaway. Ethical sourcing of Cobalt remains a concern.\\
\cline{2-8}

&
Lithium manganese oxide (LMO) &
10 kW range &
100–150  &
100 ms &
95 - 98 &
Limited & 
High power but less capacity; safer than Li-cobalt; commonly mixed with NMC to improve performance.\\
\cline{2-8}

 &
Lithium iron phosphate (LFP) &
1 kW - 50 MW &
160 - 200 &
100 ms &
93 - 98 &
Good & 
Low relative cost: \$40-60/kWh, Absence of cobalt eliminates ethical sourcing concern \\
\cline{2-8}

 &
Lithium nickel manganese cobalt oxides (LNMC) &
1 kW - 25 MW &
160 - 270 &
100 ms &
90 - 95 &
Moderate & 
High relative cost: \$420/ kWh \\
\cline{2-8}
 
 &
Lithium titanate oxide (LTO) &
1 kW - 10 MW &
200 - 260 &
100 ms &
85 - 90 &
Good & 
Considered one of the safest lithium battery chemistries available \\

\hline
Superconducting magnetic energy storage
(SMES) & 
 - &
 1 - 100 MW &
 0.5 - 5 &
 \textless 5 - 10 ms &
 95 - 99 &
 Excellent &
Offers near-immediate power injection or absorption \\
\hline

Power electronics-based supercapacitors (PESC) &
- &
 100 kW - 4 MW &
 Less than 0.2 Wh/L &
 \textless 0.1 - 1 ms &
 95 - 99 &
 Excellent &
Offers near-immediate power injection or absorption \\
\hline

\end{tabular}
\\[3pt]
\footnotesize\textsuperscript{*} Other flow battery chemistries include zinc-bromine, iron-chromium, and polysulfide-bromine.
\end{sidewaystable*}


\section{Modeling of GFLI\MakeLowercase{s} and GFMI\MakeLowercase{s}}\label{sec:modelling_of_GFM_and_GFL}

GFLI and GFMI are the two dominant categories of converter-interfaced renewable energy resources and battery energy storage systems (BESSs) in modern AI data centers. Although their hardware architectures are essentially identical, the two categories exhibit fundamentally different control philosophies and stability characteristics. On one hand, GFLIs operate as current-controlled sources that require an existing grid voltage for synchronization, typically achieved via a phase-locked loop (PLL). On the other hand, GFMIs behave as controlled voltage sources that establish local voltage and frequency references at the point of common coupling (PCC). The former primarily enables maximum renewable energy utilization by injecting available power into a stiff grid, whereas the latter is designed to provide intrinsic voltage–frequency support and enhance system stability as a primary objective. Consequently, due to these fundamental differences in dynamic behavior and control objectives, a rigorous understanding of both GFL and GFM technologies, as well as their interaction mechanisms, is essential for the design and operation of BESS-based AI data center energy supply systems. Accordingly, the following two subsections present detailed dynamic models and control structures of representative GFLIs and GFMIs considered in this study.

\subsection{Modeling of GFLIs}
Fig.~\ref{fig_Detailed_BD_GCI}(a) illustrates the general structure of an IBR that can operate either in grid-following mode, as shown in Fig.~\ref{fig_Detailed_BD_GCI}(b), or in grid-forming mode, as shown in Fig.~\ref{fig_Detailed_BD_GCI}(c). Fig.~\ref{fig_Detailed_BD_GCI}(b) shows the control structure of a grid-following inverter (GFLI) equipped with an Synchronous Reference Frame-Phase Locked Loop
(SRF-PLL) and outer power control loops, as considered in this study. The converter is supplied by a stiff DC-link voltage, typically provided by a renewable energy source or energy storage system, and is interfaced with the ac grid through an $LC$ filter composed of the filter inductance $L_\mathrm{f}$, capacitance $C_\mathrm{f}$, and equivalent series resistance $R_\mathrm{f}$. The output filter attenuates high-frequency switching harmonics and ensures smooth current injection into the grid.

Unlike grid-forming converters, the GFLI operates as a current-controlled source that synchronizes to the grid voltage using a synchronous reference frame phase-locked loop (SRF-PLL). The SRF-PLL estimates the grid voltage angle $\theta$ and angular frequency $\omega$ by forcing the $q$-axis component of the PCC voltage to zero, thereby achieving phase synchronization with the grid.

As shown in Fig.~\ref{fig_Detailed_BD_GCI}(b), the control architecture of the GFLI consists of an outer power control loop, an SRF-PLL synchronization unit, and an inner current control loop. The outer loop regulates the active and reactive power injection at the point of common coupling (PCC), denoted by $P_\mathrm{pcc}$ and $Q_\mathrm{pcc}$, respectively. Based on the tracking errors with respect to their references $P_\mathrm{ref}$ and $Q_\mathrm{ref}$, the controller generates the reference current components $i_\mathrm{d-ref}$ and $i_\mathrm{q-ref}$ in the synchronous reference frame.

The outer power control loop is typically implemented using the instantaneous power expressions in the $dq$ frame:
\begin{align}
P_\mathrm{pcc} &= \frac{3}{2}\left(v_\mathrm{d} i_\mathrm{d} + v_\mathrm{q} i_\mathrm{q}\right), \\
Q_\mathrm{pcc} &= \frac{3}{2}\left(v_\mathrm{q} i_\mathrm{d} - v_\mathrm{d} i_\mathrm{q}\right),
\end{align}
where $v_\mathrm{d}$ and $v_\mathrm{q}$ are the PCC voltage components in the synchronous reference frame.

Assuming the SRF-PLL aligns the reference frame with the grid voltage such that $v_\mathrm{q} \approx 0$, the power expressions simplify to:
\begin{align}
P_\mathrm{pcc} &\approx \frac{3}{2} v_\mathrm{d} i_\mathrm{d}, \\
Q_\mathrm{pcc} &\approx -\frac{3}{2} v_\mathrm{d} i_\mathrm{q}.
\end{align}

Therefore, the reference currents generated by the outer power control loop are obtained through PI regulators acting on the active and reactive power errors:
\begin{align}
i_{\mathrm{d,ref}} &= K_{pp}\left(P_\mathrm{ref} - P_\mathrm{pcc}\right)
+ K_{ip} \int \left(P_\mathrm{ref} - P_\mathrm{pcc}\right)\, dt, \\
i_{\mathrm{q,ref}} &= K_{pq}\left(Q_\mathrm{ref} - Q_\mathrm{pcc}\right)
+ K_{iq} \int \left(Q_\mathrm{ref} - Q_\mathrm{pcc}\right)\, dt.
\end{align}
where a unified PI structure is adopted such that $K_{pp}=K_{pq}$ and $K_{ip}=K_{iq}$, resulting in identical controller gains for both active and reactive power regulation.

The SRF-PLL plays a crucial role in ensuring accurate synchronization with the grid by estimating the phase angle $\theta_\mathrm{pll}$ using a PI-regulated loop that forces the $q$-axis voltage component to zero:
\begin{align}
\theta_\mathrm{pll} &= \int \omega_\mathrm{pll} \, dt, \\
\omega_\mathrm{pll} &= \omega_\mathrm{nom} + K_p v_\mathrm{q} + K_i \int v_\mathrm{q} \, dt.
\end{align}

The inner current control loop regulates the converter-side currents $i_\mathrm{d}$ and $i_\mathrm{q}$ to track their respective references $i_{\mathrm{d,ref}}$ and $i_{\mathrm{q,ref}}$. It is implemented in the synchronous reference frame using two identical PI controllers with decoupling and cross-coupling compensation terms, thereby improving transient response and ensuring closed-loop stability.

Overall, the GFLI operates as a grid-synchronized current source where active and reactive power injections are indirectly regulated via current references generated by the outer power control loop, while synchronization is ensured by the SRF-PLL. Therefore, in this study, a grid-forming source (either the GFMI or a gas turbine-based synchronous generator) is required to establish the PCC voltage, which serves as the synchronization reference for the SRF-PLL of the GFLI.

\subsection{Modeling of GFMI\MakeLowercase{s}}
Fig.~\ref{fig_Detailed_BD_GCI}(c) illustrates the generalized structure of the grid-forming inverter (GFMI) considered in this study. The inverter is supplied by a stiff DC voltage source representing a battery energy storage system. The converter is interfaced with the microgrid (or utility grid) through an $LC$ output filter composed of the filter inductance $L_\mathrm{f}$ and capacitance $C_\mathrm{f}$, where $R_\mathrm{f}$ denotes the equivalent series resistance of the filter inductor. The output filter effectively attenuates high-frequency switching harmonics in the inverter output voltage and current waveforms.

As shown in Fig.~\ref{fig_Detailed_BD_GCI}, the GFMI employs a cascaded control structure consisting of an inner current control loop, a voltage regulation loop, a virtual impedance loop, and a primary control layer. The inner loops regulate the converter-side currents and the point of common coupling (PCC) voltage across the filter inductance and capacitance, respectively. The virtual impedance loop enhances system damping and shapes the inverter output impedance. The primary control layer governs the synchronization dynamics and generates the voltage references required for grid-forming operation.

The primary control layer generates the internal voltage angle, $\theta=\omega t$, and the PCC voltage reference, $v_\mathrm{pcc}$. It comprises two coordinated control channels. The first channel, referred to as the active power controller (APC), regulates the active power injection at the PCC, $P_\mathrm{pcc}$, to track its reference value $P_\mathrm{ref}$ by modulating the converter internal frequency.

In the APC, several grid-forming control strategies can be implemented, including droop control, virtual synchronous generator (VSG) control, adaptive VSG (AVSG) control, virtual synchronous machine (VSM) control, matching control, and dispatchable virtual oscillator control (dVOC). In this work, droop control is adopted due to its structural simplicity, ease of implementation, and proven stability in islanded and weak grid-connected microgrid operation \cite{tayyebi2020frequency, mohammed2024grid, mohammed2025comparative}.

The GFMI operates based on the conventional $P$--$\omega$ droop mechanism, which emulates the steady-state frequency–active power characteristic of a synchronous generator by enforcing a proportional relationship between active power imbalance and frequency deviation.

The converter operating frequency is given by \cite{guerrero2010hierarchical}:
\begin{align}
\omega &= \omega_\mathrm{nom} - m_p \left(P_\mathrm{ref} - P_\mathrm{pcc}\right), \\
\theta &= \int \omega \, dt,
\end{align}
where $\omega_\mathrm{nom}$ is the nominal angular frequency, $P_\mathrm{pcc}$ and $P_\mathrm{ref}$ denote the measured and reference active powers, respectively, and $m_p$ is the active power–frequency droop coefficient.

The droop coefficient is defined as \cite{mohammed2022accurate}:
\begin{align}
m_p = \frac{2\pi\left(f_\mathrm{max} - f_\mathrm{min}\right)}{2 P_\mathrm{rated}},
\end{align}
where $P_\mathrm{rated}$ is the rated active power of the converter, while $f_\mathrm{max}$ and $f_\mathrm{min}$ define the admissible frequency limits.

The second control channel regulates the converter voltage magnitude. In conventional grid-forming implementations, a reactive power controller (RPC) is typically used, where reactive power injection $Q_\mathrm{pcc}$ is regulated via voltage magnitude modulation.

In contrast, this work adopts a $Q$--$V$ droop-based voltage regulation strategy. The $d$-axis voltage reference is generated as:
\begin{align}
V_\mathrm{d,ref} &= V_\mathrm{nom} + n_q \left(Q_\mathrm{ref} - Q_\mathrm{pcc}\right),
\end{align}
where $V_\mathrm{nom}$ denotes the nominal voltage magnitude, $Q_\mathrm{pcc}$ is the measured reactive power at the PCC, and $Q_\mathrm{ref}$ is its reference value. The droop gain $n_q$ is defined as:
\begin{align}
n_q = \frac{V_\mathrm{max} - V_\mathrm{min}}{2 Q_\mathrm{rated}}.
\label{eq_4}
\end{align}

Here, $Q_\mathrm{rated}$ represents the rated reactive power capability of the GFMI. The factor of two accounts for the bidirectional reactive power exchange capability. The parameters $V_\mathrm{max}$ and $V_\mathrm{min}$ define the permissible voltage deviation limits in accordance with grid code requirements.

By enforcing $v_{\mathrm{q,ref}} = 0$, the voltage magnitude satisfies $|v_\mathrm{dq}| \approx v_\mathrm{d}$ under balanced steady-state conditions. Consequently, the voltage regulation loop ensures accurate PCC voltage tracking, eliminates steady-state voltage deviation, and indirectly governs reactive power exchange with the grid.

\begin{figure*}[!t]
  \centering
  \subfloat[General structure of an inverter-based resource (IBR) that can operate either in grid-following mode in grid-forming mode\label{fig2_a}]{%
     \includegraphics[width=0.7\linewidth]{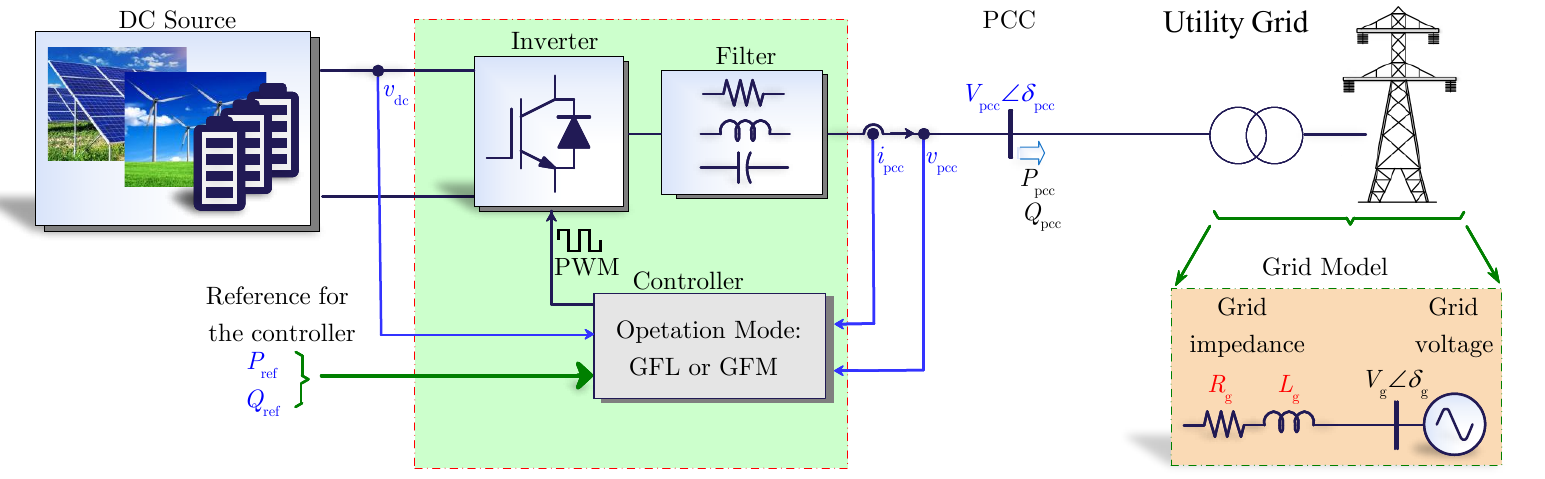}
   } %
   \hfill
  \subfloat[Control structure of the GFLI equipped with an SRF-PLL and outer power control loops\label{fig_BD_GCI}]{%
     \includegraphics[width=0.7\linewidth]{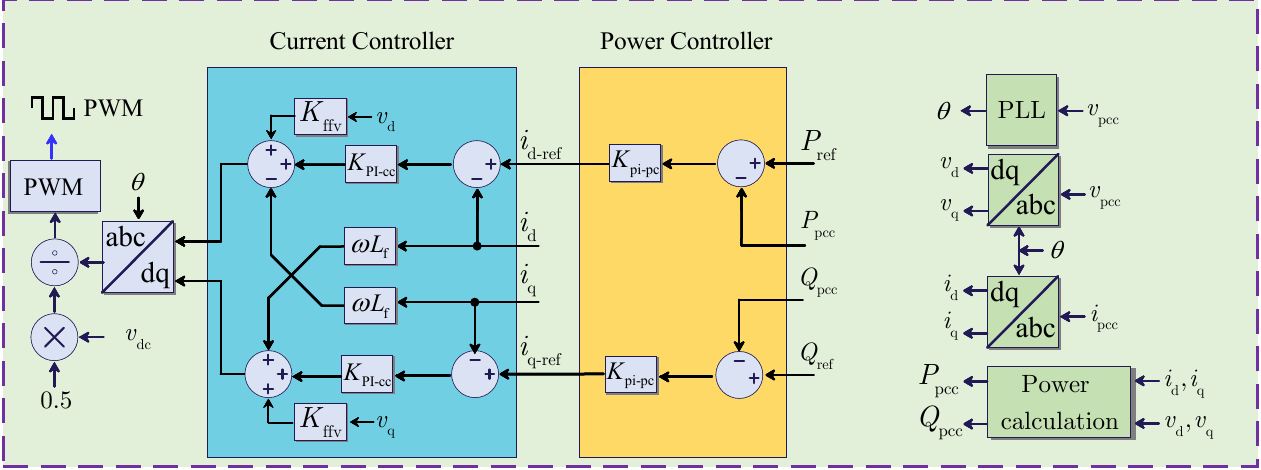}
   } %
   \hfill
  \subfloat[Control structure of the GFMI equipped with droop-based primary control\label{fig_Detailed_BD_GCI}]{%
      \includegraphics[width=0.7\linewidth]{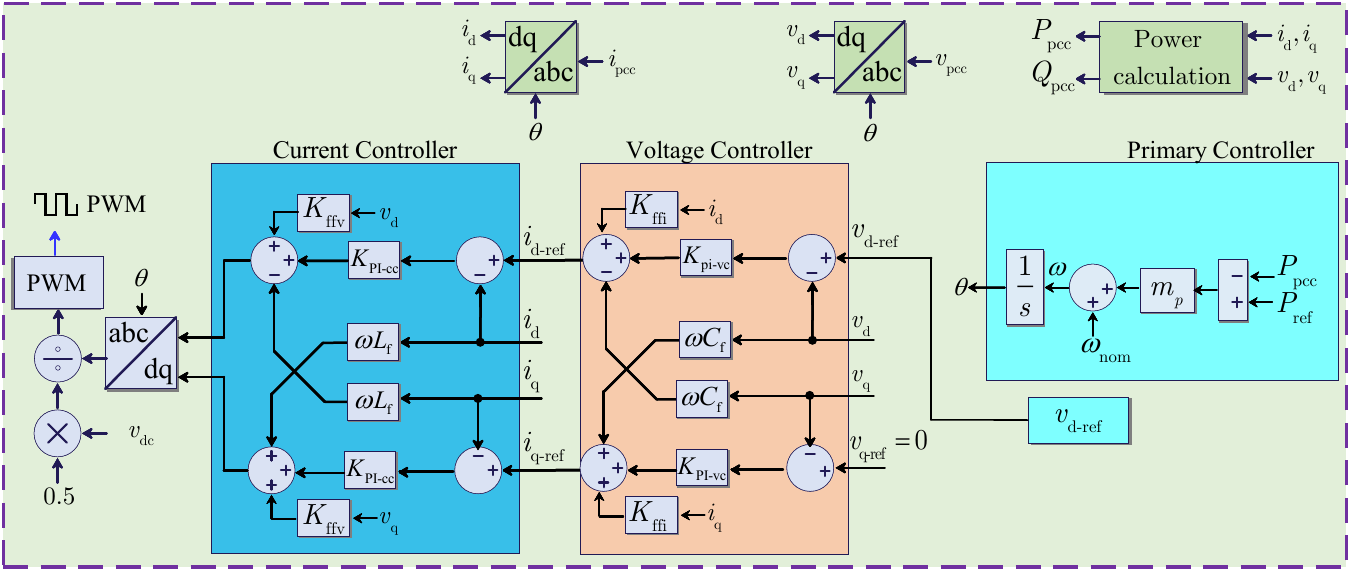}
  }
\caption{
Electric circuit and control structures of the investigated IBRs}
  \label{fig_Detailed_BD_GCI} 
\end{figure*}

\section{Results and discussion} \label{sec_simulation_results}
In this section, electromagnetic transient (EMT) simulation results with different energy system combinations are evaluated. The overall data center plant configuration was derived from Fig. \ref{fig:phased_approach_a}. The AI load profile from Fig. \ref{fig:Train_ft_inference} was emulated and scaled to represent a data center plant load of 300 MW; this arrangement is intended to be an aggregate representation of a typical modern hyperscaler. A single generation source or a combination of sources was used to demonstrate how these sources can cater to the dynamic load behavior of the AI data center. 





Seven cases were investigated, with Table~\ref{tab:summary_of_result}  providing a concise overview of these cases along with recommendations on preferred and non-preferred configurations. Key parameters used for the simulation are documented in Table ~\ref{tab:parameters}.

\begin{table*}[t]
\small
\centering
\caption{Summary of simulation cases from section \S\ref{sec_simulation_results} and observations}
\label{tab:summary_of_result}
\renewcommand{\arraystretch}{1.2} 
\begin{tabular}{p{1.25cm} p{1.25cm} p{3cm} p{4cm} p{5cm}}
\hline
\textbf{Case number} &
\textbf{Data center load}  &
\textbf{Mode of deployment} &
\textbf{Generation composition for on-site generation (rated value)*} &
\textbf{Observations}\\
\hline
Case 1 &
300 MW &
Islanded generation &
Gas turbine plants: 10 × 40 MVA units (represents an N+2 case) & Base power generating source usually unable to handle AI load spikes, causing large swings in system frequency. Profile not recommended for off-grid AI data center application.\\
\hline
Case 2 &
300 MW &
Islanded generation &
Gas turbine plants: 9 × 40 MVA units, GFL BESS-PV system: 50 MVA,  & GFL BESS follows the system dynamics set by the on-site gas turbine units. Large system swings persist. Profile not recommended for off-grid AI data center application. \\
\hline
Case 3 &
300 MW &
Islanded generation &
GFM BESS-PV system: 200 MVA, Gas turbine plants: 4 × 40 MVA units & Fast-acting GFM BESS-PV system tackles the AI load spikes, and the gas turbines adjust slowly. Profile recommended for off-grid AI data center application.\\
\hline
Case 4 &
300 MW &
Islanded generation &
GFM BESS-PV system: 350 MVA &  GFM BESS-PV acts as an islanded microgrid and maintains frequency and voltage within tight threshold limits.\\
\hline
Case 5 &
300 MW &
Islanded generation &
GFM BESS-PV system: 300 MVA, 
GFL BESS-PV system: 50 MVA &  
(a) GFM BESS-PV system dominates the system and maintains frequency and voltage within tight threshold limits. 

(b) GFL BESS-PV system provides a stable 50 MVA output.

(c) Profile recommended for off-grid AI data center application.
 \\

\hline
Case 6 &
300 MW &
Fully grid-connected deployment transitioning to islanded generation due to grid  disturbances &
GFM BESS-PV system: 300 MVA, Gas turbine plants: 2 × 40 MVA units &  (a) Final built represents a higher percent of cleaner GFM BESS-PV system. 
(b) Millisecond-level transients exist, but stable on-site power is quickly achieved.

(c) Profile recommended for AI data center application that may not ride through during grid-induced disturbances.\\

\hline
Case 7 &
N/A &
Restoration of data center with power grid &
N/A &
(a) Droop-based GFM provides baseline stable operation but limited inertia support during resynchronization.

(b) VSG-based GFM improves transient performance via virtual inertia but exhibit sensitivity to grid impedance and SCR variations.

(c) AVSG-based GFM provides robust performance, ensuring improved damping, faster resynchronization, and stable operation across weak-to-strong grid transitions, making it the most suitable profile for AI data center restoration scenarios.
\\

\hline
\end{tabular}
\\[3pt]
\footnotesize\textsuperscript{*} Represents the maximum rated value of the generating source. Generating sources are usually operated at less than their maximum operational limits to accommodate load variability and growth margin.
\end{table*}


\begin{table}[t]
\centering
\caption{Key parameters used for simulation study}
\label{tab:parameters}
\renewcommand{\arraystretch}{1.2} 
\begin{tabular}{p{1.5cm} p{1cm} p{1cm} p{3cm} }
\hline
\multicolumn{4}{c}{\textbf{System parameters}}\\
\hline
Quantity &
Value &
Unit &
Description \\

\hline
\textit{f} &
60 &
Hz & 
Nominal frequency\\

\textit{V} &
13.2 &
kV & 
Nominal voltage of the MV bus with 480 V to 13.2 kV inverter step-up transformer\\

\hline
\multicolumn{4}{c}{\textbf{Gas turbine parameters}}\\
\hline
$S_{rated}$ &
40 &
MVA &
Apparent power capacity \\

\hline
$X_d,X_q$ &
1.9, 1.8 &
p.u. & 
Direct and quadrature axis synchronous reactances\\

$X'_d,X'_q$ &
0.15, 0.3 &
p.u. & 
Direct and quadrature axis transient reactances\\

$X''_d,X''_q$ &
0.13, 0.21 &
p.u. & 
Direct and quadrature axis subtransient reactances\\
\hline

\multicolumn{4}{c}{\textbf{Parameters for single grid-following inverter (Total are 300 units)}}\\
\hline
                $v_\mathrm{t}$  &  480 &  V  &  Terminal voltage (L-L, RMS)\\
                $P_\mathrm{rated}$  &  1.0 & MW  &  Inverter rated power\\
                $v_\mathrm{dc}$  &  1700  &V  &  DC Bus voltage\\
                $f_\mathrm{sw}$  &  5 &kHz  &  Switching frequency\\
                $L_\mathrm{f}$  &  95& $\mu$H  &  Filter inductance\\
                $R_\mathrm{f}$  &  10&  m$\Omega$  &  Filter resistance\\
                \hline

\multicolumn{4}{c}{\textbf{Parameters for single grid-forming inverter (Total are 300 units)}}\\
\hline
                $v_\mathrm{t}$  &  480 &  V  &  Terminal voltage (L-L, RMS)\\
                $P_\mathrm{rated}$  &  1.0 & MW  &  Inverter rated power\\
                $v_\mathrm{dc}$  &  1700  &V  &  DC Bus voltage\\
                $f_\mathrm{sw}$  &  5 &kHz  &  Switching frequency\\
                $L_\mathrm{f}$  &  95& $\mu$H  &  Filter inductance\\
                $R_\mathrm{f}$  &  10&  m$\Omega$  &  Filter resistance\\
                $C_\mathrm{f}$  &  2&  mF  &  Filter capacitance\\
                $m_p$  &  3.14$\times 10^{-6}$ & rad/W.s &  $P-\omega$ Droop coefficient of the APC\\
                $n_\mathrm{q}$  &  $55.2\times10^{-6}$  & V/Var &  $Q-V$ Droop coefficient of the RPC\\
                \hline
\hline

\end{tabular}
\end{table}

\subsection{Case 1 - Gas turbine only}\label{Case 1}
\begin{figure*}[htbp]
    \centering

    \subfloat[Cumulative data center load profile - typical of AI training \label{fig_case1a}]{%
        \includegraphics[width=0.65\linewidth]{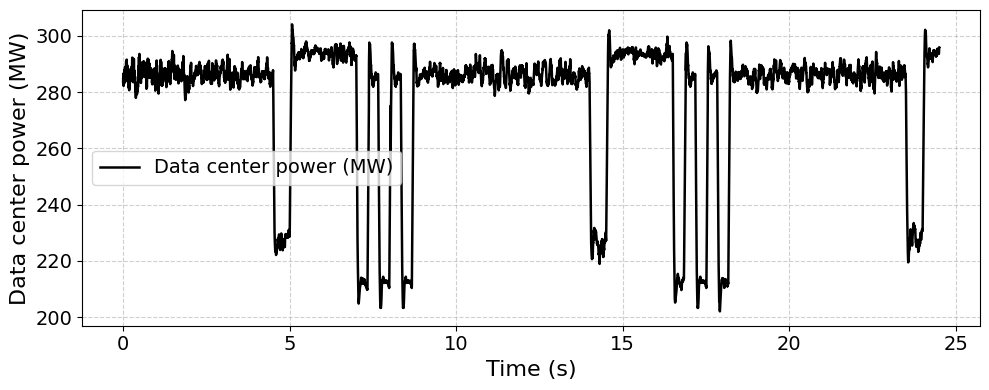}
    }
    \hfill

    \subfloat[Plot showing generation from on-site gas turbines \label{fig_case1b}]{%
        \includegraphics[width=0.65\linewidth]{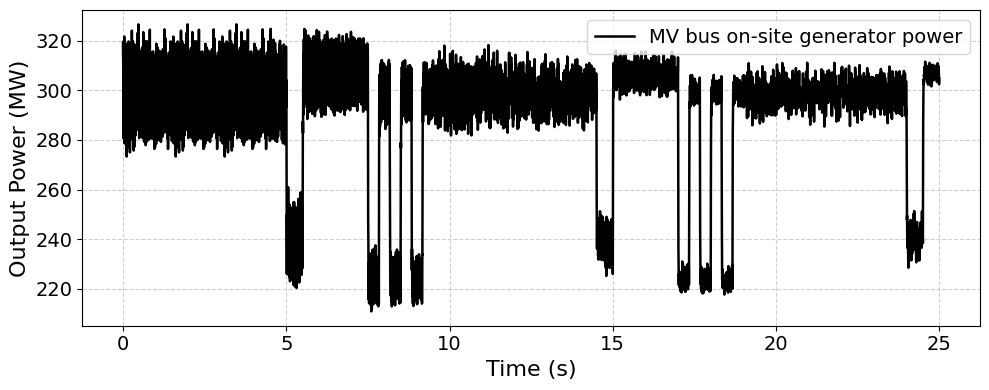}
    }
    \hfill
    
     \subfloat[Voltage at the data center substation's MV bus\label{fig_case1c}]{%
        \includegraphics[width=0.65\linewidth]{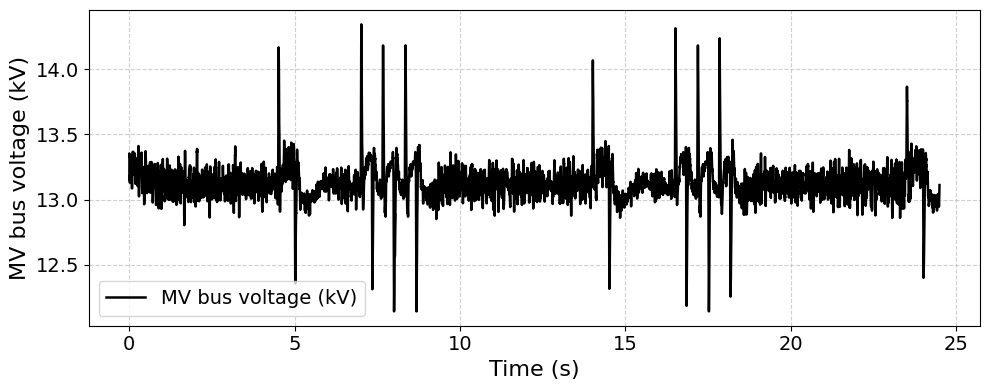}
    }
    \hfill
    \subfloat[Frequency at the data center substation's MV bus\label{fig_case1d}]{%
        \includegraphics[width=0.65\linewidth]{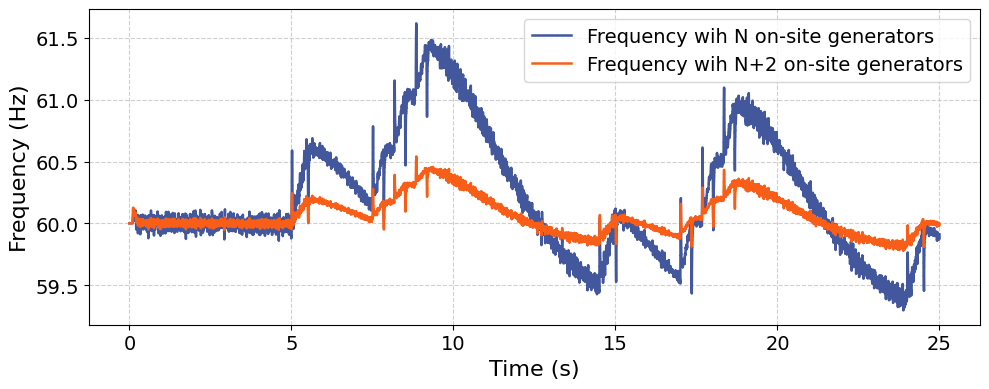}
    }

    \caption{Illustration of case 1 - Operation of the data center plant with on-site gas turbines generation.}
    \label{fig:case1}
\end{figure*}

This first case in this series consists exclusively of gas turbine plants, each rated 40 MVA. The output of the gas turbines is fed into a 13.2 kV MV bus to which AI data center load feeds are connected. These data centers are assumed to be operated by a single tenant (which prevents diversification of the AI load profile and hence represents a conservative case), with the cumulative data center power draw shown in Fig.~\ref{fig_case1a}. This cumulative data center power draw is derived directly from the AI training load profile, as previously shown in Fig.~\ref{fig:Train_ft_inference}.\\
Fig.~\ref{fig_case1b} shows the generation from the on-site gas turbines. From the on-site gas turbine generation, one may observe that the output of the on-site gas turbine generators mirrors the AI load profile. Fig.~\ref{fig_case1c} shows instantaneous voltage spikes that the on-site gas turbines are able to ride through. Fig.~\ref{fig_case1d} presents the output frequency with 8 and 10 units of 40 MVA gas turbine generators, respectively. As can be seen, 8 - 40 MVA gas turbines are able to arithmetically meet the maximum data center load (referred to as N on-site generators in Fig.~\ref{fig_case1d}), but the frequency excursions are prolonged and reach up to 61.5 Hz. Adding two additional generators (referred to as N+2 on-site generators in Fig.~\ref{fig_case1d}) allows improvement in the plant's frequency response.\\
It should, however, be noted that adding additional generation units to improve the plant's overall frequency response comes at the cost of additional capital expenditure (CAPEX) and, overall, may be difficult to implement from an air quality permitting standpoint in certain jurisdictions. Additionally, there exist concerns about these frequency excursions and the corresponding cumulative and excessive turbine torques, decreasing the overall life of the on-site generator to a significant extent.   \\
In general, unless aggressive software-level mitigation or GPU-level power-smoothing techniques are employed \cite{choukse2025power, zhao2023sustainable, maheshwari2026ai} (or hybridized energy systems exist as shall be discussed in this section later), standalone on-site gas turbines are unsuitable to serve AI loads due to the risk of turbine damage from these load variations. One important aspect to note here is that for new greenfield phased deployments of non-AI data centers, such as enterprise, cloud, or edge data centers (such as those used for extending rural network streaming or 5G capability) with a relatively flat load profile, standalone on-site gas turbine-based generators may have leverage. \\

\subsection{Case 2 - Gas turbine and GFL BESS-PV system} \label{Case 2}
This case represents an extension of Case 1, with some portion of the generation MVA being provided by an on-site GFL BESS-PV system. GFL system, by its design, follows the dynamics of the dominant source, the gas turbine generation units. The power system dynamics and the deployment limitations of this implementation remain similar to those of Case 1 and Fig.~\ref{fig:case1}. The incremental advantage of the on-site GFL BESS-PV system is that it may help relieve some of the air quality permitting limits associated with the gas turbine generation units if those limits are reached or maxed out.

\subsection{Case 3 - Gas turbine and GFM BESS-PV system} \label{Case 3}
\begin{figure*}[htbp]
    \centering


    \subfloat[Plot showing generation from GFM BESS-PV system and on-site gas turbines \label{fig_case3b}]{%
        \includegraphics[width=0.65\linewidth]{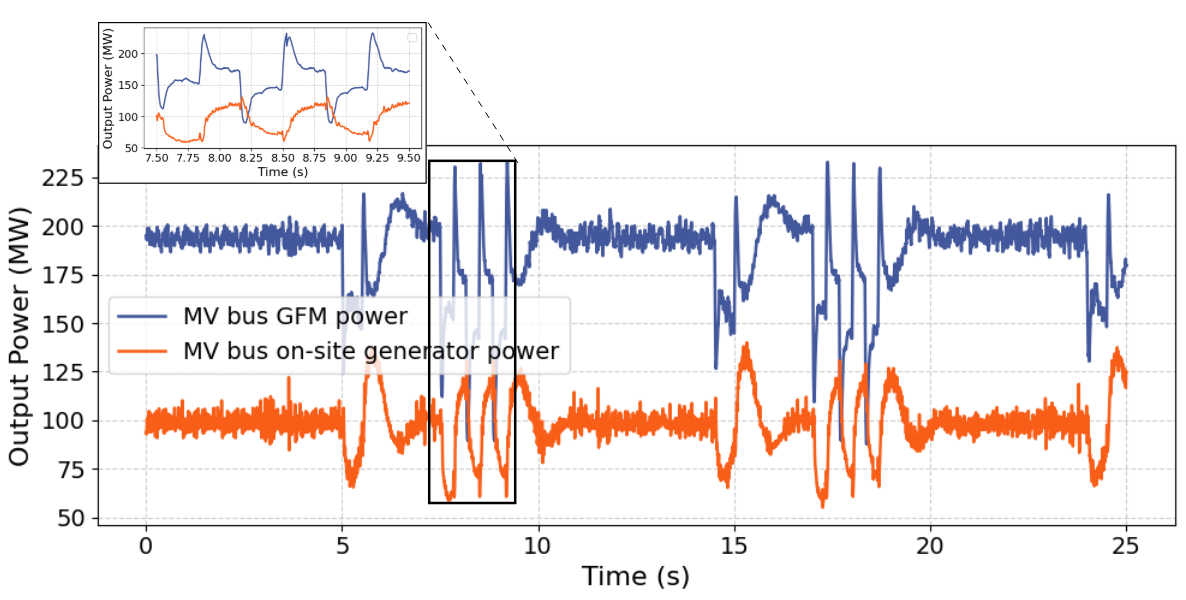}
    }
    \hfill
    
     \subfloat[Voltage at the data center substation's MV bus\label{fig_case3c}]{%
        \includegraphics[width=0.65\linewidth]{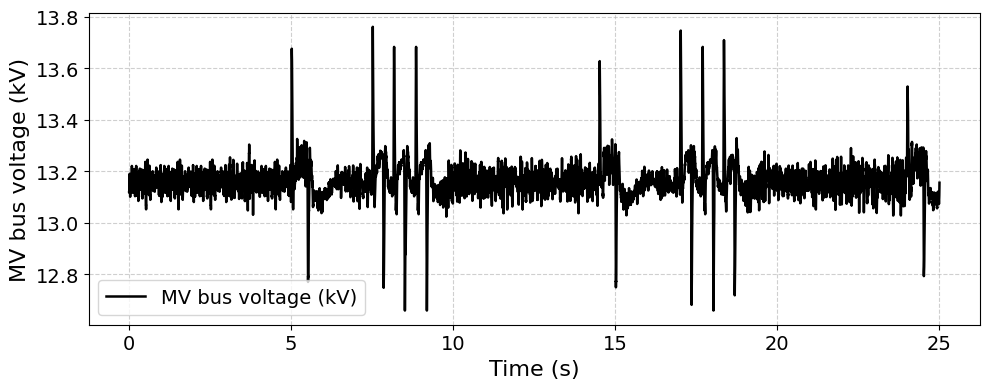}
    }
    \hfill
    \subfloat[Frequency at the data center substation's MV bus\label{fig_case3d}]{%
        \includegraphics[width=0.65\linewidth]{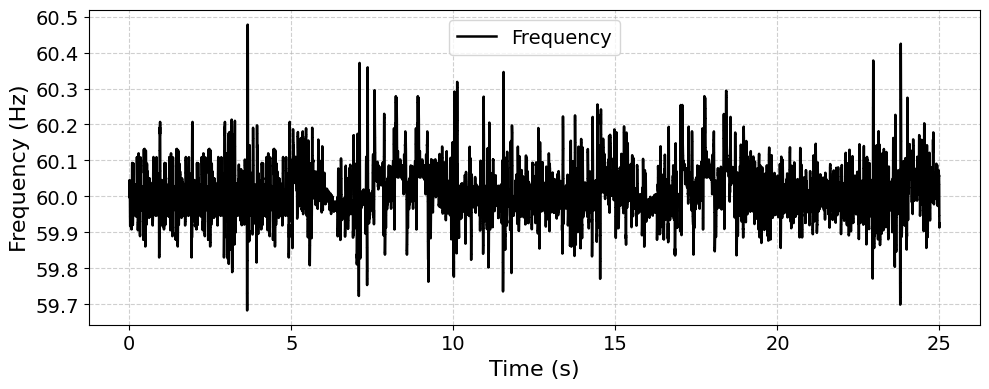}
    }

    \caption{Illustration of case 3 - Operation of the data center plant with GFM BESS-PV system and on-site gas turbines generation.}
    \label{fig:case3}
\end{figure*}

Simulations are continued with the same AI training load profile for the data center plant as before; Fig.~\ref{fig_case1a}, with a cumulative data center plant load of 300 MW. The generation system consists of 4 × 40 MVA gas turbine generators and a 200 MVA GFM BESS-PV system. The GFM BESS-PV system dominates the islanded system and allows for rapid change in its output power. As one may observe from Fig.~\ref{fig_case3b}, with the fluctuation of the training load profile between 300 and 200 MW, the GFM BESS-PV system modulates its output power rapidly and works in synergy with the slower responding on-site gas turbine generators to maintain the MV bus voltage and the frequency to remain within certain thresholds. \\
The high-frequency, low-magnitude spikes observed in the MV bus voltage and frequency profiles, Fig.~\ref{fig_case3c}, \ref{fig_case3d}, remain within a threshold of 0.95 - 1.05 p.u., and are expected and common with AI training load profiles. Depending on the data center's power distribution architecture, the input power is generally conditioned/ filtered at the distribution level double-conversion UPS, or at the rack-level PDUs.

\subsection{Case 4 - GFM BESS-PV system only} \label{Case 4}
\begin{figure*}[htbp]
    \centering


    \subfloat[Plot showing generation from GFM BESS-PV systems \label{fig_case4b}]{%
        \includegraphics[width=0.65\linewidth]{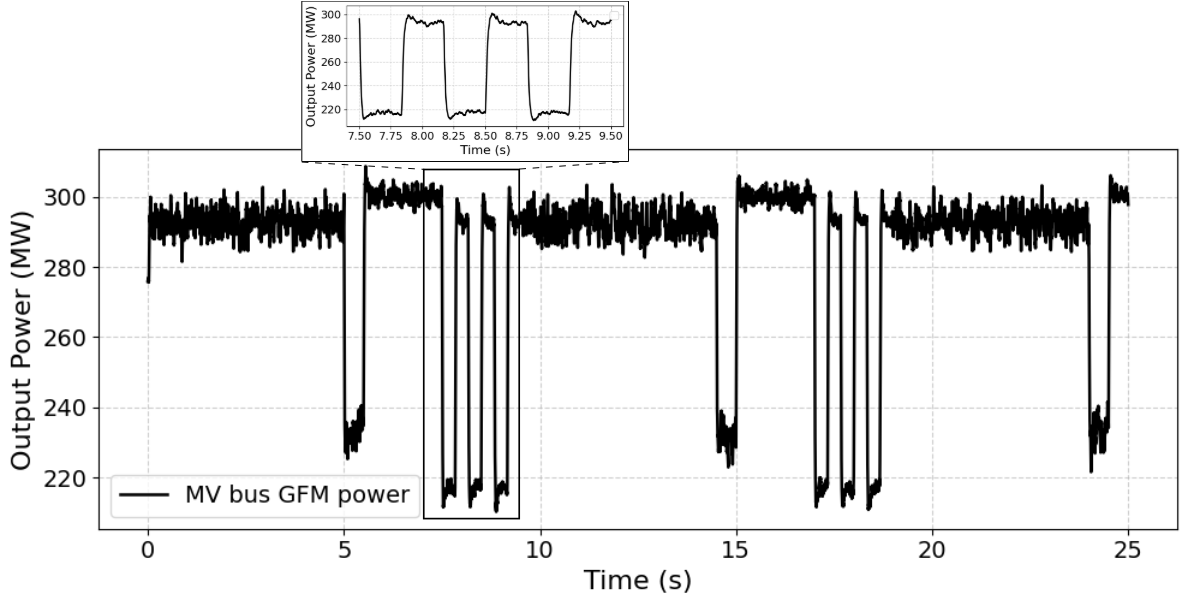}
    }
    \hfill
    
     \subfloat[Voltage at the data center substation's MV bus\label{fig_case4c}]{%
        \includegraphics[width=0.65\linewidth]{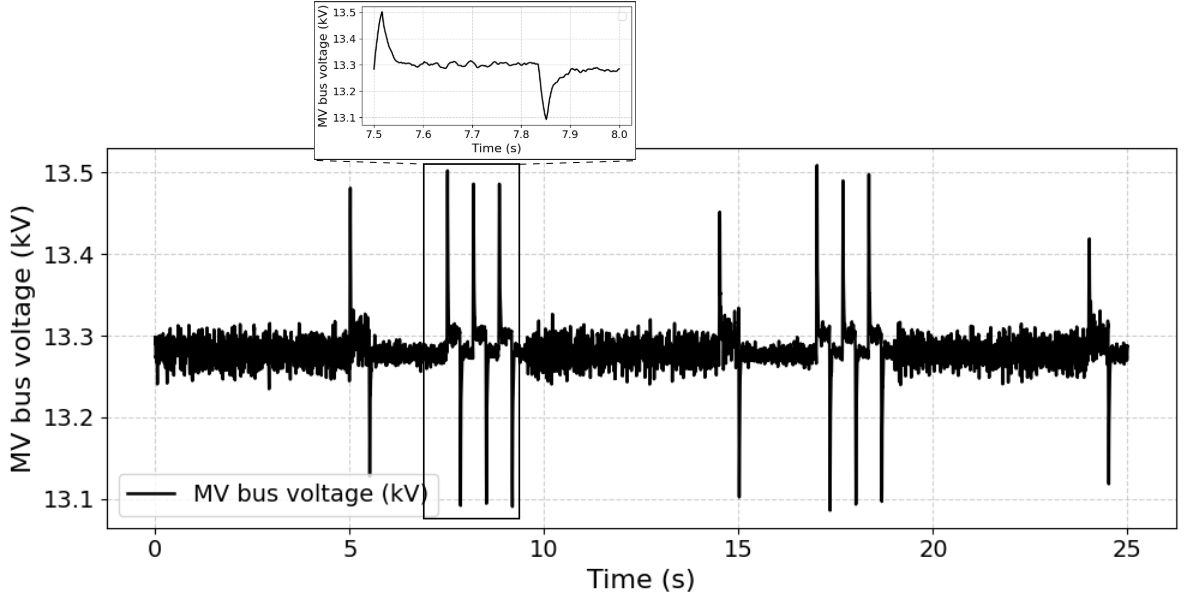}
    }
    \hfill
    \subfloat[Frequency at the data center substation's MV bus\label{fig_case4d}]{%
        \includegraphics[width=0.65\linewidth]{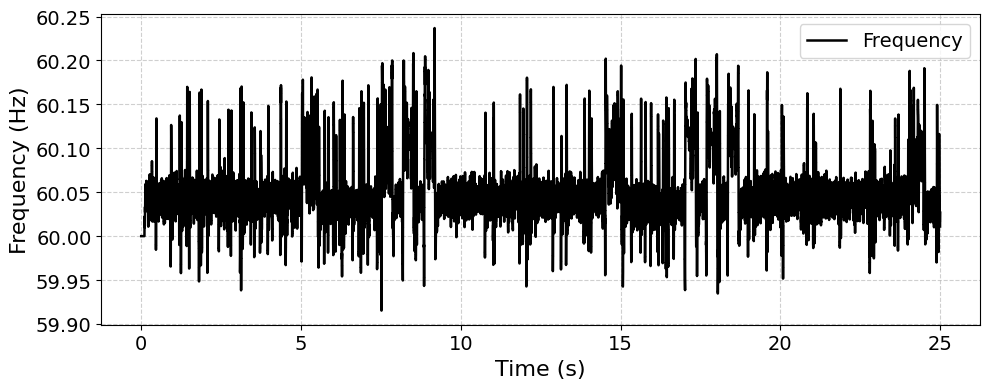}
    }

    \caption{Illustration of case 4 - Operation of the data center plant with GFM BESS-PV systems.}
    \label{fig:case4}
\end{figure*}

For this next case, the same AI training load profile (Fig.~\ref{fig_case1a})
for the data center plant is maintained. The generation system consists exclusively of GFM BESS-PV system. Such a system may be leveraged in areas where there might be very little emission latitude, making deployment of natural gas turbines difficult or infeasible from an air-quality permitting perspective. With 'time-to-deployment' a primary driver for data center developers and given the fairly short lead times for utility scale LFP battery systems, an islanded GFM BESS-PV system emerges as a practical option.\\
As observed from the power output plot, Fig. Fig.~\ref{fig_case4b}, the GFM controller exhibits fast and precise dynamic response and is able to mimic the AI load profile by modulating its output. The MV bus voltage, as shown in Fig.~\ref{fig_case4c}, remains within tight margins of 0.95 to 1.05 pu (13.1 to 13.5 kV), with 13.2 kV as the nominal value. Similarly, the MV bus frequency, as shown in Fig.~\ref{fig_case4d}, remains within a tight threshold of 59.9 and 60.25 Hz, further illustrating the practicality of such a system.

\subsection{Case 5 - GFM and GFL BESS-PV system  } \label{Case 5}
\begin{figure*}[htbp]
    \centering


    \subfloat[Plot showing generation from GFM and GFL BESS-PV systems \label{fig_case5b}]{%
        \includegraphics[width=0.65\linewidth]{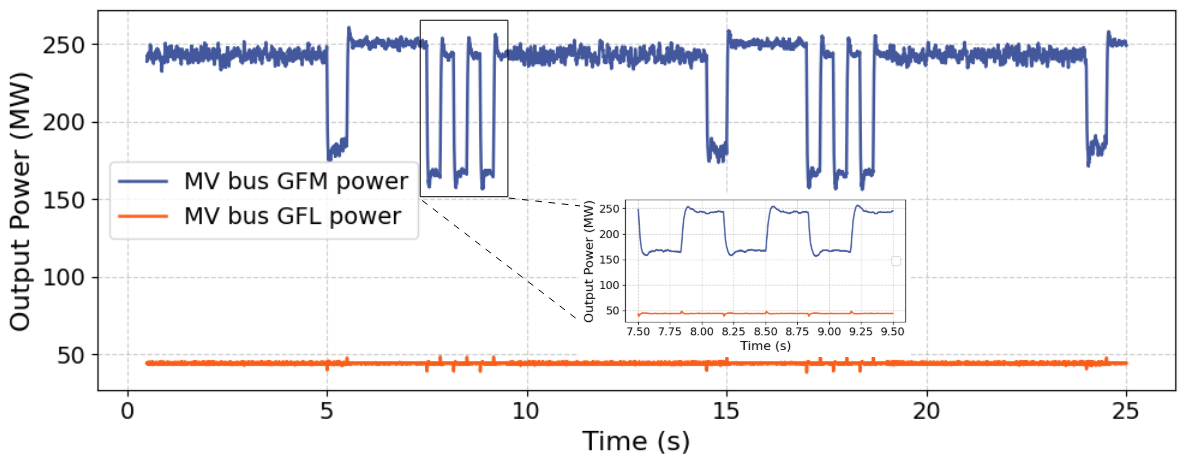}
    }
    \hfill
    
     \subfloat[Voltage at the data center substation's MV bus\label{fig_case5c}]{%
        \includegraphics[width=0.65\linewidth]{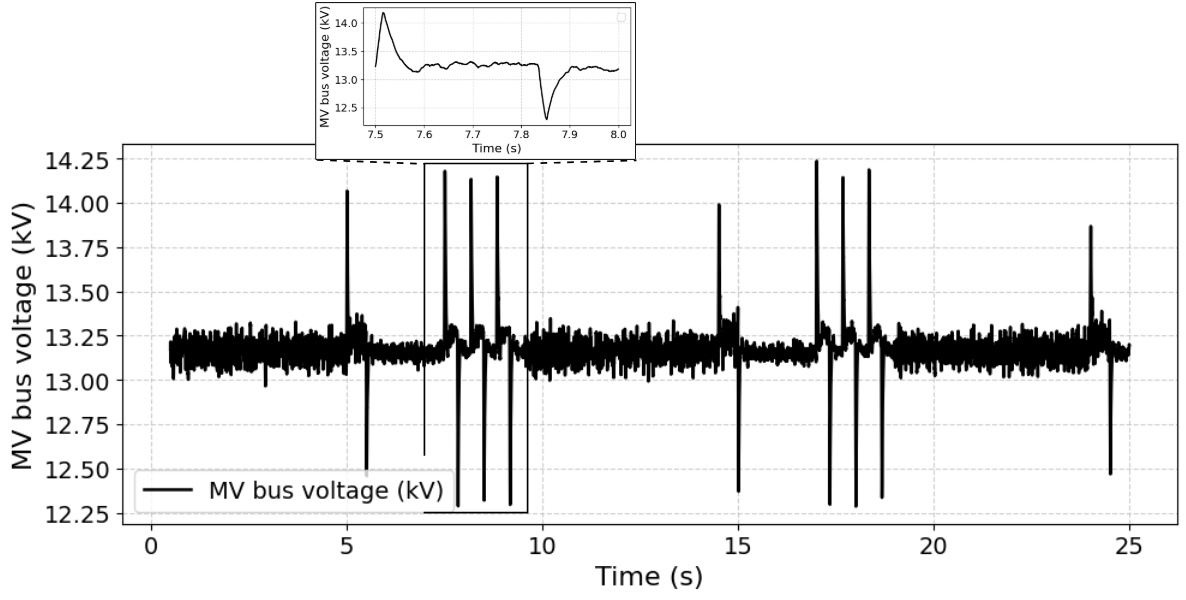}
    }
    \hfill
    \subfloat[Frequency at the data center substation's MV bus\label{fig_case5d}]{%
        \includegraphics[width=0.65\linewidth]{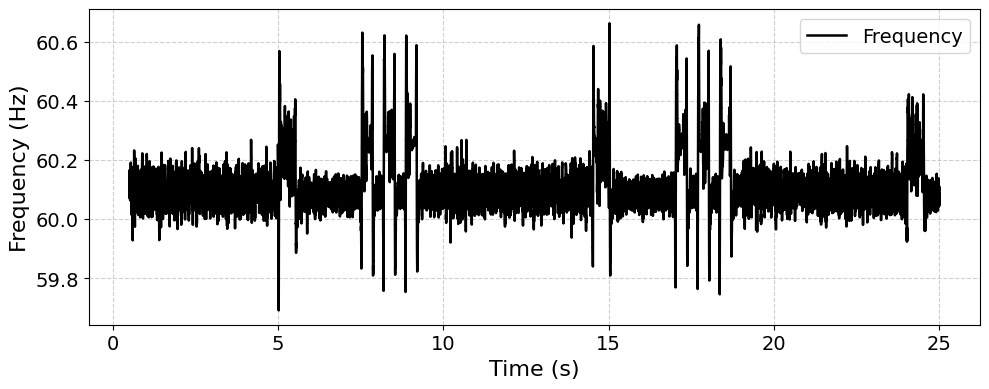}
    }

    \caption{Illustration of case 5 - Operation of the data center plant with GFM and GFL BESS-PV systems.}
    \label{fig:case5}
\end{figure*}

This particular case represents a scenario where a new data center facility is sited and developed near an existing PV plant, thereby leveraging the benefit of co-location. In such colocation cases, the existing PV plant is usually expected to have traditional GFL-based IBRs. For this illustration, such a 50 MW GFL-based PV plant is integrated with the data center facility, which houses an additional  300 MW of GFM BESS-PV system. A similar set of simulations is conducted, with the data center load profile, the results of power generated by GFM and GFL BESS-PV systems, the system voltage, and the system frequency plotted in Fig.~\ref{fig_case5b} through \ref{fig_case5d}. As expected, the GFM BESS-PV system dominates, setting the voltage and frequency of the islanded grid, while the GFL BESS-PV system provides a stable target power output set at 50 MVA. The GFM BESS-PV system rapidly modulates its power output in response to fluctuations in the AI training load. Similar to the previous cases, these high-frequency, low-magnitude spikes are generally conditioned/ filtered at the distribution level double-conversion UPS, or at the rack-level PDUs, before being fed into the GPUs.

\subsection{Case 6 - Fully grid-connected operation and transition to islanded operation} \label{Case 6}
\begin{figure*}[htbp]
    \centering

    \subfloat[Plot showing loss of grid power with GFM BESS-PV system and on-site standby gas turbines picking up the plant load\label{fig_case6a}]{%
        \includegraphics[width=0.65\linewidth]{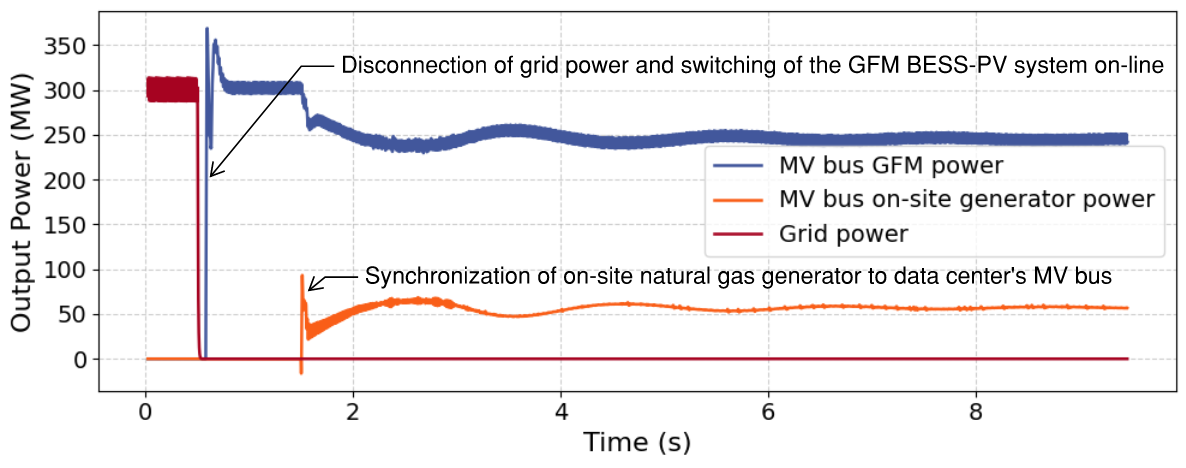}
    }
    \hfill
    
     \subfloat[Voltage at the data center substation's MV bus\label{fig_case6b}]{%
        \includegraphics[width=0.65\linewidth]{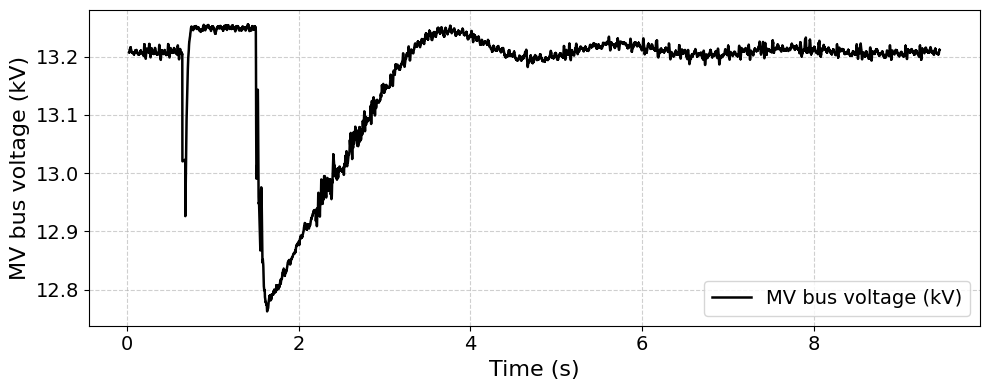}
    }
    \hfill
    \subfloat[Frequency at the data center substation's MV bus\label{fig_case6c}]{%
        \includegraphics[width=0.65\linewidth]{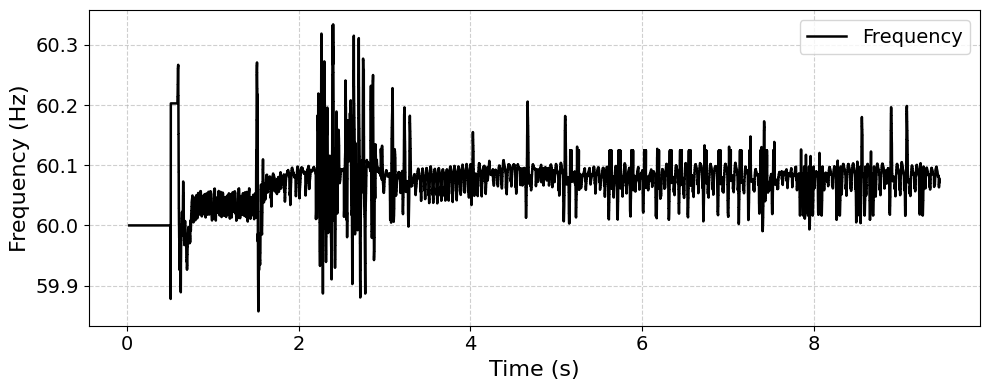}
    }

    \caption{Illustration of case 6 - Operation of the data center plant in grid connected mode and transition into islanded on-site BESS and natural gas generation.}
    \label{fig:case6}
\end{figure*}

The cases being studied shall be those of a fully grid-connected operation, where the data center plant is being run with grid power. This represents a fully executed interconnection agreement maturing to commercial operation, via the interconnecting substation, usually at 345 or 138 kV. For this case, dynamics of the system for a grid to on-site transition shall be of interest. There are enough documented events of such grid to on-site transition, primarily in the wake of voltage disturbances on the grid that cause the data center plant to transition to an islanded mode of operation, rather than riding through the event. In one particular incident \cite{NERC_sensitive_load} from July 10, 2024, from the US state of Virginia, a shunted surge arrestor on a 230 kV transmission line triggered multi-shot automatic recloser operation, causing six recordable voltage depressions, from back-to-back reclose operation at the two ends of the transmission line. The event triggered customer-driven disconnection of roughly 1,500 MW of data center load, which transitioned to on-site power. \\
For this case, the interest lies in the grid to on-site transition, and not the AI load ramp behaviors, and hence a flat data center load profile was simulated. At 0.5 s the grid power into the 300 MW data center plant was disconnected, and the on-site GFM BESS-PV system comes on-line to pick up the plant load. At the same time, a start and synchronization signal was received by the small fleet of two 40 MVA gas turbine plants, which were brought online at 1.5 s into the simulation. Fig.~\ref{fig_case6a} showcases the disconnection of 300 MW of grid power and picking up of the plant load by a combination of on-site GFM BESS-PV system and on-site natural gas generator power. Fig.~\ref{fig_case6b} illustrates the momentary dip in voltage to 12.9 kV at the 13.2 kV bus, during the grid to on-site GFM BESS-PV system transition. This voltage dip may be managed by reducing the delay associated with the communication and control system of the GFM BESS-PV. A second voltage dip of 12.8 kV is witnessed upon the synchronization of the on-site natural gas generator, originating primarily due to the interaction of the GFM BESS-PV system, which ramps down its output to accommodate the output power from the natural gas generating unit. Fig.~\ref{fig_case6c} shows the frequency dynamics, with momentary frequency spikes, reaching upto 60.3 Hz, observed during the disconnection of grid power and during the synchronization of the on-site natural-gas generator.

\subsection{Case 7 - Restoration of AI data center back onto the grid - the need for adaptive GFM control} \label{Case 7}
The cases presented so far in this study consider droop-based grid-forming (GFM) control as the baseline approach, owing to its simplicity and robust steady-state power–frequency characteristics, making it the most widely adopted implementation in practical systems.

AI data centers impose stringent performance requirements on grid-interfacing converters due to the high sensitivity of large-scale computational loads to voltage and frequency disturbances during both normal operation and restoration events. In particular, aggregated server power supplies require tightly regulated point-of-common-coupling (PCC) voltage, limited frequency excursions, and rapid transient recovery to maintain stable operation and avoid load shedding, service interruptions, or computational instability. During restoration scenarios, when AI data center loads are resynchronized and reconnected to the utility grid, the interfacing converters are expected to provide frequency support, controlled active-power ramping, and adequate damping while maintaining stable operation across a broad range of grid conditions, extending from weak distribution networks with low short-circuit ratios (SCRs) to stiff transmission-level systems.

These operational requirements highlight the limitations of conventional droop-based GFM control and motivate the investigation of more advanced GFM strategies. While droop control provides effective power-sharing characteristics, it lacks inertia emulation capability and may therefore be insufficient for applications requiring enhanced frequency support and improved transient performance during restoration events. To assess the suitability of alternative GFM approaches for AI data center restoration, a comparative time-domain analysis of four representative GFM control strategies is presented.

\begin{figure*}[!t]
    \centering

    \subfloat[PCC active power response.]{
        \includegraphics[width=0.35\linewidth]{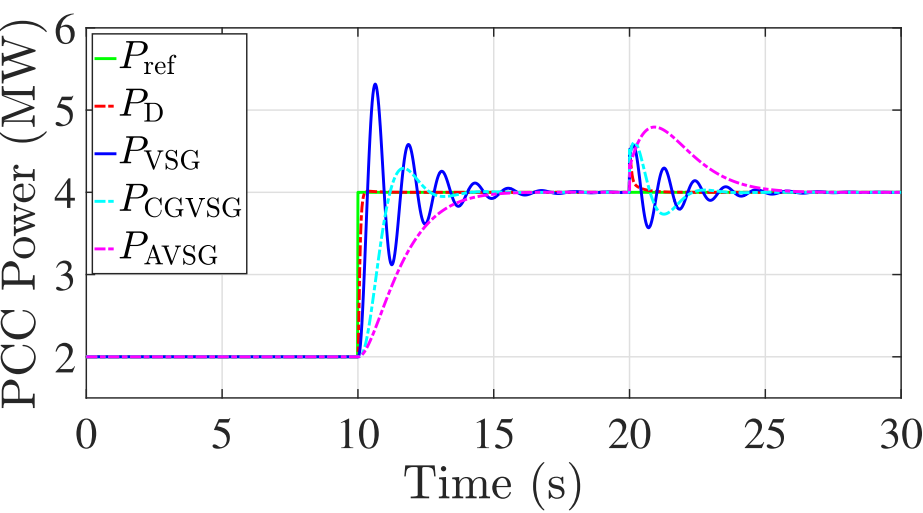}
        \label{fig:Pref_step_SCR8_XR7_DVCA}
    }
    \hfill
    \subfloat[PCC reactive power response.]{
        \includegraphics[width=0.35\linewidth]{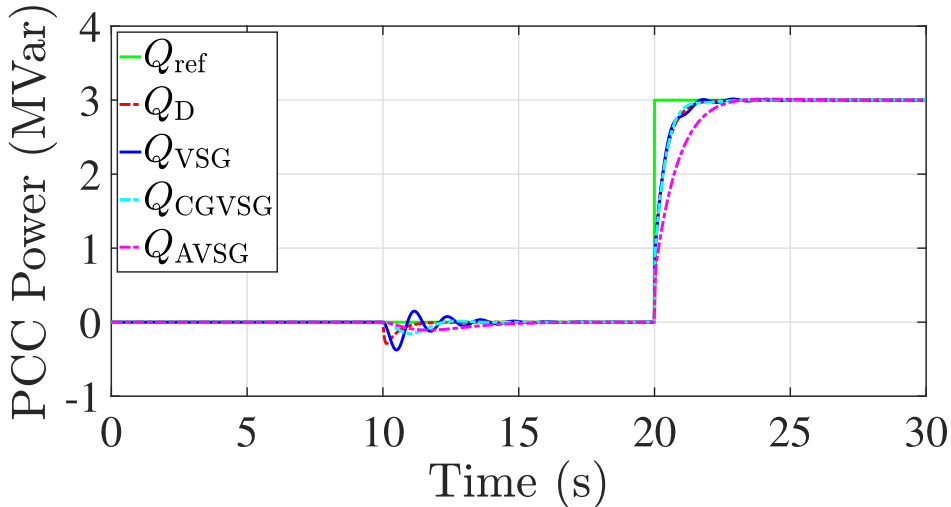}
        \label{fig:Qref_step_SCR8_XR7_DVCA}
    }
    \hfill
    \subfloat[GFMI frequency response.]{
        \includegraphics[width=0.35\linewidth]{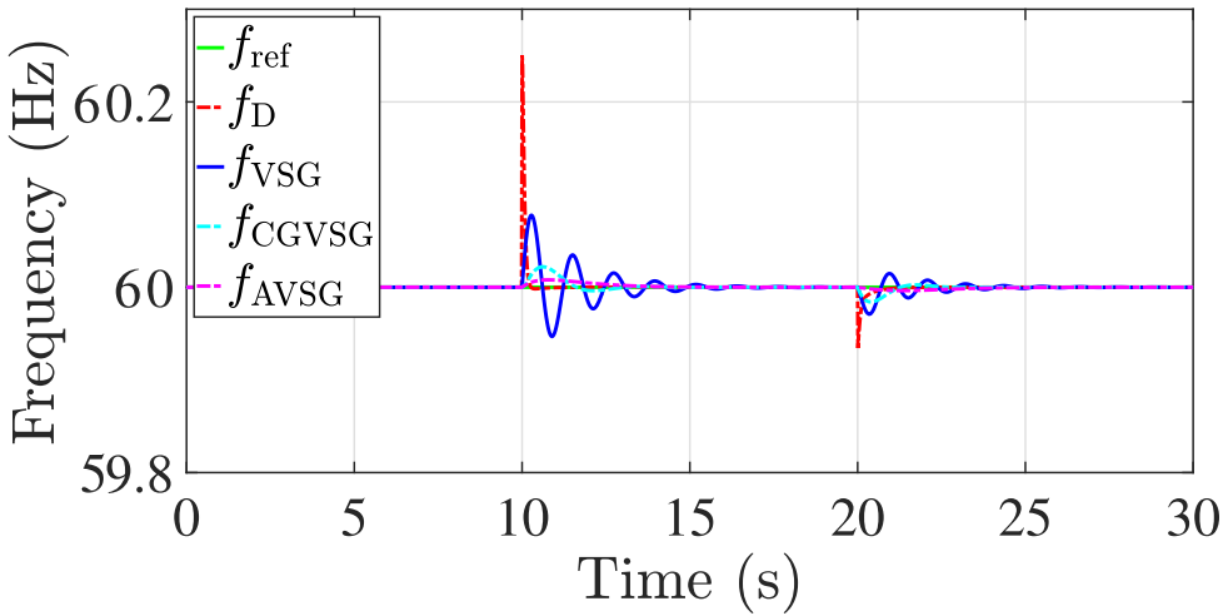}
        \label{fig:Freq_step_SCR8_XR7_DVCA}
    }

    \caption{Performance of the GFMIs in a strong grid with SCR = 8 and $X_g/R_g = 7$ under active- and reactive-power reference step changes. The active-power reference, $P_{\mathrm{ref}}$, is increased from 2 MW to 4 MW (0.5 to 1.0 p.u.) at $t = 10$ s, while the reactive-power reference, $Q_{\mathrm{ref}}$, is increased from 0 MVAr to 3 MVAr (0 to 1.0 p.u.) at $t = 20$ s. 
    \cite{mohammed2024grid}.
    }
    \label{fig_Pref_Qref_step_SCR8_XR7_DVCA}
\end{figure*}

Fig.~\ref{fig_Pref_Qref_step_SCR8_XR7_DVCA} illustrates the dynamic performance of the GFMIs operating in a strong-grid environment characterized by an SCR of 8 and an $X_g/R_g$ ratio of 7 under active- and reactive-power reference step changes. At $t=10$ s, the active-power reference is increased from 2 MW to 4 MW (0.5 to 1.0 p.u.), followed by a reactive-power reference step from 0 MVAr to 3 MVAr (0 to 1.0 p.u.) at $t=20$ s. The results show that the conventional droop-based controller exhibits the fastest reference-tracking response and the shortest settling time among the evaluated methods. However, this fast dynamic response is achieved without the provision of virtual inertia, thereby limiting its contribution to frequency support and inertial response during transient disturbances.

In contrast, the virtual synchronous generator (VSG) controller introduces synthetic inertia but exhibits pronounced oscillatory behavior and reduced damping, resulting in larger transient deviations and longer settling times. This observation is consistent with the known sensitivity of VSG-based control strategies to variations in the Thevenin-equivalent grid impedance, particularly under grid-connected operation, where changes in grid strength and network characteristics can adversely affect damping performance and dynamic stability \cite{mohammed2022online}. In strong-grid conditions, the PCC voltage is largely dictated by the upstream network, reducing the converter's effective voltage regulation authority and potentially degrading power decoupling and damping characteristics.

The CSGVS and adaptive virtual synchronous generator (AVSG) controllers achieve a more favorable trade-off between response speed, damping performance, and stability robustness. Among the investigated strategies, the AVSG controller demonstrates the best overall dynamic performance, characterized by smooth active- and reactive-power responses, enhanced damping, negligible overshoot, and well-regulated frequency dynamics. The superior performance of the AVSG is attributed to its adaptive control framework, which employs online grid impedance estimation and continuously updates controller parameters according to prevailing grid conditions \cite{mohammed2022online}. By adapting to variations in the Thevenin-equivalent grid impedance, the AVSG improves power decoupling, mitigates low-frequency oscillations, and maintains consistent transient performance across a wide range of grid strengths and impedance characteristics.

These attributes are particularly advantageous for AI data center restoration applications, where stable resynchronization, secure grid reconnection, and robust operation under dynamically varying network conditions are essential for maintaining power quality and ensuring uninterrupted computational services.

To further assess the performance metrics of the considered GFM techniques, Table~\ref{tab:grid-forming-compare} presents a comparative evaluation of representative GFM control strategies in AI data center restoration scenarios. In this analysis, the GFM inverters are assumed to operate on an islanded bus and remain connected throughout the grid restoration and resynchronization process. The table highlights key differences among the evaluated control strategies in terms of transient response characteristics, damping performance, and sensitivity to variations in SCR. Overall, the comparison emphasizes the inherent trade-offs associated with different GFM control implementations under varying grid conditions and operating regimes. For a more detailed characterization of the underlying GFM techniques, including both frequency-domain and time-domain analyses, the interested reader is referred to the comprehensive studies reported in \cite{mohammed2024grid, mohammed2025comparative}.

\begin{table*}[!t]
\renewcommand{\arraystretch}{1.25}

\caption{Performance comparison of grid-forming control strategies for AI data center reconnection and restoration under varying conditions.}
\label{tab:grid-forming-compare}
\centering

\resizebox{1.95\columnwidth}{!}{
\begin{tabular}{p{4.5cm} | c | c | c | c | c | c}
\hline\hline \\[-3mm]

\textbf{Performance Criteria} &
\textbf{Droop-based GFM} &
\textbf{VSG-based GFM} &
\textbf{CGVSG-based GFM} &
\textbf{AVSG-based GFM} &
\textbf{Matching-based GFM} &
\textbf{dVOC-based GFM} \\ \hline

Data center load reconnection stability (robustness to grid SCR increases) &
\cmark\cmark & 
\cmark & 
\cmark\cmark\cmark & 
\cmark\cmark\cmark & 
\cmark & 
\cmark\cmark \\ \hline

Response of IT load demand (operating point variations) &
\cmark\cmark & 
\cmark & 
\cmark\cmark\cmark & 
\cmark\cmark\cmark & 
\cmark & 
\cmark\cmark \\ \hline

Synthetic inertia capability &
\xmark & 
\cmark\cmark & 
\cmark\cmark\cmark & 
\cmark\cmark\cmark & 
\xmark & 
\xmark \\ \hline

PCC voltage support for IT loads &
\cmark\cmark\cmark & 
\cmark\cmark\cmark & 
\cmark\cmark\cmark & 
\cmark\cmark\cmark & 
\cmark\cmark\cmark & 
\cmark\cmark\cmark \\ \hline

Sensitivity against frequency deviations &
\cmark\cmark\cmark & 
\cmark\cmark\cmark & 
\cmark\cmark\cmark & 
\xmark &
\cmark\cmark\cmark & 
\cmark\cmark\cmark \\ \hline

Oscillation suppression during reconnection &
\cmark\cmark & 
\cmark & 
\cmark\cmark\cmark & 
\cmark\cmark\cmark & 
\cmark & 
\cmark\cmark \\ \hline

Robustness against $X_\mathrm{g}/R_\mathrm{g}$ increases &
\cmark\cmark & 
\cmark & 
\cmark\cmark\cmark & 
\cmark\cmark\cmark & 
\cmark & 
\cmark\cmark \\ \hline

Robustness against symmetrical voltage sag &
\cmark\cmark & 
\cmark\cmark & 
\cmark\cmark & 
\cmark\cmark\cmark & 
\cmark\cmark & 
\cmark\cmark \\ \hline

Requirement of $R_\mathrm{g}$ and $X_\mathrm{g}$ knowledge in tuning &
No & 
No  & 
Yes & 
Yes & 
No & 
No \\ \hline

Implementation complexity &
Low & 
Moderate  & 
Moderate to High & 
Moderate & 
Moderate & 
Moderate to High \\ \hline

\end{tabular}
}
\\[3pt]
\footnotesize Levels of satisfaction of the controller performance: \cmark\cmark\cmark\ very good, \cmark\cmark\ good, \cmark\ acceptable, \xmark\ poor (or unaccepted).
\end{table*}



\section{Conclusion and Future Recommendations} \label{sec_conclusion}
This work develops a detailed framework for phased development of AI data centers, emphasizing interconnection constraints associated with integrating very large, highly concentrated loads into the bulk power system and showing that AI training demand profiles are intricate and fundamentally distinct from conventional utility loads. The pronounced ramping behavior characteristic of AI training can be partially mitigated through GPU- or software-level control; however, significant challenges remain in tracking these rapidly varying power trajectories in real time at the facility level. A systematic comparison of common energy storage technologies suitable for AI data centers, including multiple lithium-ion chemistries, indicates that lithium iron phosphate (LFP) batteries offer the most favorable overall balance of performance, safety, and lifecycle considerations for this application.

The proposed three-layered, phased development strategy takes into account multi-year execution timelines for grid interconnection agreements and practical procurement lead times for major equipment, such as power transformers and high-voltage circuit breakers. The development framework is structured so that the early and intermediate stages rely on islanded operation enabled by a hierarchical on-site energy architecture. In these stages, on-site natural-gas turbines coordinated with a grid-forming BESS–PV microgrid provide sufficient flexibility to accommodate the fast and irregular AI load fluctuations while operating decoupled from the transmission network. In the final development phase, the site transitions toward a configuration in which on-site standby generation and a grid-forming BESS–PV system support continued operation in the event of grid disconnection or severe grid disturbances. Multiple grid-forming inverter control options are evaluated in the context of AI data center loading, and an adaptive virtual synchronous generator–based control scheme is recommended because of its reduced sensitivity to short-circuit strength variations and its improved damping of transient oscillations under changing grid conditions.

As large AI data centers become a significant share of new grid-connected loads, a coherent and forward-looking regulatory framework will be essential to preserve bulk system reliability and security. To that end, the following future recommendations are proposed: 

\subsection{Requiring more uniform adoption of ride-through capabilities} More consistent adoption of ride-through capabilities should be mandated. At present, the ride-through criteria specified by different TSOs/ISOs are heterogeneous and often incomplete, and large data centers may not be required to sustain uniform voltage and frequency ride-through performance during system disturbances. Harmonized and stringent ride-through requirements for data centers would reduce the likelihood of multiple facilities tripping offline simultaneously, thereby mitigating the risk of cascading or widespread grid disruptions.
\subsection{Requiring more uniform definition of ramp rate limits during disturbances} Ramp-rate limits for large AI loads during normal and disturbed conditions should be defined in a more standardized way. Rather than relying on project-by-project AI load ramp assumptions provided by developers during transmission interconnection studies, the interconnecting utility or TSO/ISO should specify standardized ramp-rate envelopes for both normal operation and contingency scenarios. Such a framework would reduce unexpected power swings, lessen the need to oversize generation and storage assets, and facilitate more consistent assessment of system impacts.
\subsection{Standardizing large load plant model validation requirements} Standardized plant model validation requirements for large loads should be extended and formalized. Forward-looking guidance is needed to ensure that detailed EMT models of data centers are anchored to measured plant-level performance, analogous to the mature validation practices used for renewable generation plants. These existing renewable integration standards can serve as a template for developing large-load model validation protocols. In addition, Hardware-in-the-Loop (HIL)–based approaches should be incorporated into the design and validation of control systems associated with AI data center loads, variable frequency drives, inverter-based resources, PV systems, and BESS installations connected to the bulk power grid, thereby enabling more realistic testing of control interactions under a wide range of operating conditions. \\

By adopting a phased development strategy, data center developers can maintain an edge on their "time-to-market" deployment goals, decreasing some of the dependence on fulfillment of grid interconnection agreements. 

\section*{Generative AI Usage Statement}
During the preparation of this work, the author(s) used Claude (Anthropic) to assist with sentence formatting and grammar checking. After using this tool, the author(s) reviewed and edited the content as needed and took full responsibility for the content of the published article.

\section*{Funding}
This research received no external funding. This work was independently conducted and self-funded by the authors.

\section*{Declaration}
The authors declare that they have no known competing financial interests that could have appeared to influence the work reported in this paper. \\
The views, thoughts, opinions, and conclusions made in this material are solely those of the authors and don’t necessarily reflect the views of the authors' employer, organization, committee, or other group or individual. 

\inConf{
\bibliographystyle{IEEEtran}
\bibliography{references}
}
\inArxiv{
\bibliographystyle{tmlr}
\bibliography{references}
}


\end{document}